\documentclass[aps,nofootinbib,showpacs,preprintnumbers,amsmath,amssymb]{revtex4}
\def\be{\begin{equation}}
\def\ee{\end{equation}}
\def\ba{\begin{eqnarray}}
\def\ea{\end{eqnarray}}
\def \bea{\begin{eqnarray}}
\def \eea{\end{eqnarray}}

\usepackage{graphics}
\usepackage{graphicx}% Include figure files
\usepackage{dcolumn}% Align table columns on decimal point
\usepackage{bm}% bold math
\usepackage{epsfig}
\usepackage{graphicx}
\usepackage{multirow}
\usepackage{dcolumn}% Align table columns on decimal point
\usepackage{graphicx,epsfig}%
\usepackage{enumerate}

\def \ee{\end{equation}}
\def \be{\begin{equation}}
\def \bea{\begin{eqnarray}}
\def \eea{\end{eqnarray}}

\preprint{}
\begin{document}

%\tableofcontents

\title{Differentiable-Path Integrals in Quantum Mechanics}

\author{ Benjamin Koch$^*$ and Ignacio Reyes$^*$}
 \affiliation{
$^*$ Instituto de F\'{i}sica, \\
Pontificia Universidad Cat\'{o}lica de Chile, \\
Av. Vicu\~{n}a Mackenna 4860, \\
Santiago, Chile \\
}
\date{\today}

\begin{abstract}
A method is presented which restricts the space of paths entering the path integral of quantum mechanics to subspaces of $C^\alpha$, by only allowing paths which possess at least $\alpha$ derivatives. The method introduces two external parameters, and induces the appearance of a particular time scale $\epsilon_D$ such that for time intervals longer than $\epsilon_D$ the model behaves as usual quantum mechanics.
However, for time scales smaller than $\epsilon_D$, modifications to standard formulation
of quantum theory occur. This restriction renders convergent some quantities which are usually divergent in the time-continuum limit $\epsilon\rightarrow 0$. We illustrate the model by computing several meaningful physical quantities such as the mean square velocity $\langle v^2 \rangle $, the canonical commutator, the Schrodinger equation and the energy levels of the harmonic oscillator. It is shown that an adequate choice of the parameters introduced makes the evolution unitary.
\end{abstract}

%\pacs{04.62.+v, 03.65.Ta}
\maketitle

%%%%%%%%%% FIGURE %%%%%%%%%%%%%%%%%%

%%%%%%%%%%%%%%%%%%%%%%%%%%%%%%%%%%%%%%%%%%%%

%%%%%%%%%%%%%%%%%%%%%%%%%%%%%%%55
\section{Introduction}\label{Introduction}

%%%%%%%%%%%%%%%%%%%%%%%%%%%%%%%55
What are the relevant trajectories in the path integral approach to quantum mechanics? The path integral (PI) method, envisioned by Dirac and developed by Feynman \cite{Fey1948, FeyHib1965} is defined by a summation over all possible histories or configurations going from an initial to a final state. For the case of quantum mechanics this means a summation
over all continuos functions connecting two events in space-time. 
However some natural questions arise at a very early stage, for example: are  all paths relevant, important, or even necessary to define a consistent quantum theory? How do different ``classes"\ of paths contribute to the path integral? What is the effect of leaving some paths out? 
Specifically, what is the role of the highly irregular/nowhere-differentiable paths in this summation?
The topic regarding this last question was already discussed in \cite{FeyHib1965}, and thereafter further developed by many others 
\cite{FeyHib1965,Rivers1987,AbbotWise1980,Zinn-Justin2005,Schulman1981,AAS1997,Nottale2008}:
\textit{``The important paths for a quantum mechanical particle are not those which have a definite slope (or velocity)... Typical paths of a quantum mechanical particle are highly irregular on a fine scale... In other words, the paths are nondifferentiable."}\cite{FeyHib1965}

%%%%%%%%%%%%%%%%%%%%%%%%%%%%%%%%%%%%%%%%%%%%%%%%%%%%%%%%%%%%%%%%%%%%%%%%%%%%%%%%%%
 \begin{figure}[hbt]
   \centering
%\centerline{\protect\vbox{\epsfig{file=Approx.eps,
%width=0.6\textwidth}}}
\includegraphics[width=10cm]{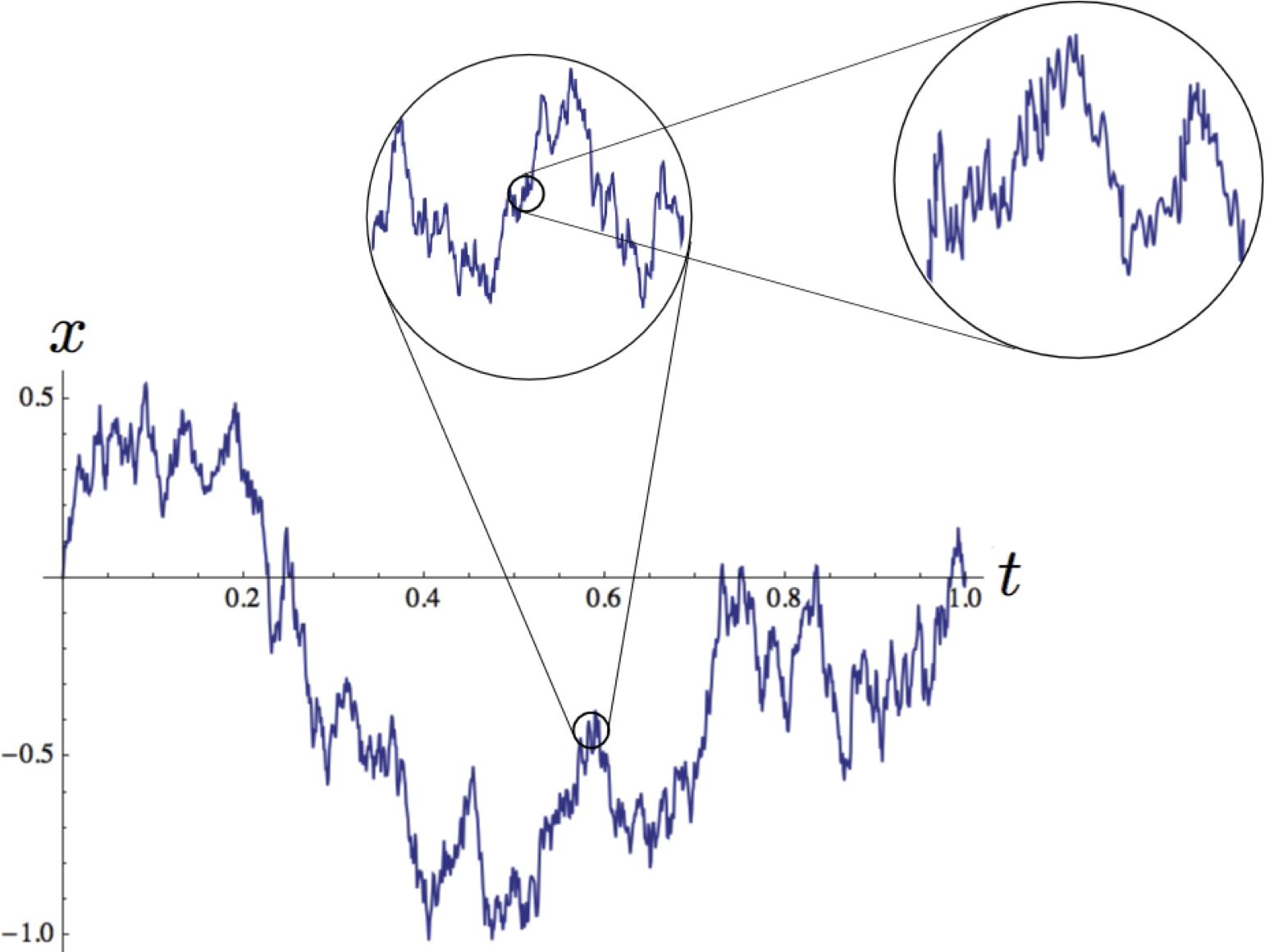}
  \caption{typical quantum trajectory in Feynman's approach: curves are continuos but nowhere-differentiable.}
\label{fig:nondiff}
\end{figure}

From this analysis there has arisen a mainstream point of view, which may be summarized in three key statements:
\begin{enumerate}[(A)]
\item The dominant contribution to the path integral comes from the nowhere-differentiable paths rather than from the differentiable ones. This is quantified in the definition of the Wiener measure,
which is $1$ for the nowhere-differentiable functions and $0$ for the differentiable ones. \cite{Szabados1994, Wiener1923,Hunt1994}

\item These nowhere-differentiable trajectories are fractal (self-similar), and their velocity diverges as $\langle v^2 \rangle \sim 1/\epsilon$, where $\epsilon $ is the time interval of measurement defining the slope, and therefore they possess infinite action. 
\item This singular behavior of nowhere-differentiable paths at small scales is indispensable for obtaining usual quantum mechanics (e.g. the commutation and uncertainty relations) 
\end{enumerate}

In this sense the time slicing parameter $\epsilon$ can be seen as a regulator, very much like the
any regulator in quantum field theory (QFT), which is known to be a priori notoriously divergent.
The difference between QM and QFT is that in the latter case almost all physical observables turn out to be divergent quantities so one is compelled to impose additional conditions (regularization - renormalization procedure) in order to obtain finite answers.
To the opinion of the authors it would be however much more satisfactory 
(bearing in mind the passage to the much more pathological QFT) to
have a meaningful limit $\epsilon \rightarrow 0$ already in QM, for divergent quantities
such as the mean squared velocity $\langle v^2 \rangle $. \\

The observations above have led to the common belief that the summation over nowhere-differentiable curves is unavoidable for formulating a coherent
quantum theory in the PI language.
However, this idea is not a logical necessity of the affirmations (A)-(C): observation (A) is only relevant
when the path integral involves summing over  both nowhere-differentiable  and
differentiable functions. 
Instead, we propose to consider a point of view which has been previously discarded: 
to restrict the space of paths, 
summing over differentiable functions only.
%\footnote{Note that the idea of restricting the space of paths over which one sums has 
%been successfully applied before: e.g. Faddeev-Popov method \cite{FadPop1967} narrows the functional space, 
%giving rise to a ``well defined" path integral which would otherwise be evidently ill defined. }. 
By doing so in a simplistic and straightforward approach, a consistent quantum theory will be constructed which recovers 
the canonical quantum theory in a broad regime of scales but differs 
from the usual quantum theory in the regime of very short times.\\
%It will be shown that the restriction to differentiable paths (here formulated within a specific model) will
%actually allow to compute those originally problematic quantities in a finite
%and well defined way, without getting into conflict with the well tested ``observable'' results
%of the usual quantum theory.\\

Because QM is non-relativistic, the issue of divergences could seem futile, since one knows the theory will break down anyway at some High Energy Physics scale. Therefore the aim of this paper is not so much to make a definite statement about QM, but rather to give a sensible method to eliminate divergences that works for QM, which then will enable us to apply it QFT, a truly relativistic ultraviolet (UV) divergent theory. We must warn that no attempt of mathematical rigor is made. Rather, we wish to assess the physical viability of constructing path integrals over differentiable-controlled paths. \\ 

The article is organized as follows: section \ref{sec:Differentiable paths} briefly presents a method for controlling the paths and how we implement it in practice; in section \ref{sec: General results} some general results are shown concerning the geometrical nature of the method and its effect on commutation and uncertainty relations; section \ref{sec:SED} shows how the wave equation is modified; section \ref{sec: Specific results} presents two particular examples (free particle and harmonic oscillator) along with dealing with the issue of Unitarity, which is supported by the numerical analysis of section \ref{sec:NumRes}. Finally in section \ref{sec: Summary, Discussion and Conclusions} it is suggested some possible connections with other approaches and pose some future research topics. For ease of comparison with conventional quantum mechanics, in each topic of Sections \ref{sec: General results}-\ref{sec: Specific results} we first present first the ``Feynman case" in which we show how to derive the standard results as a limiting case of our method, and next present the ``Differentiable case" showing the computation in the differentiable path integral model.  \\

\underline{Remark on notation}: in this paper, as will become evident, the symbol $\langle F(t) \rangle$ (for some $F$) represents a path integral-statistical average, and not an inner product in Hilbert space associate to a specific wave function
\begin{align}\label{}
\langle F (t) \rangle=\frac{\int \mathcal{D}x\ e^{-\frac{1}{\hbar}S[x(t)]} F(t) }{\int \mathcal{D}x\ e^{-\frac{1}{\hbar}S[x(t)]}} \neq \langle \psi | \hat F | \psi \rangle
\end{align}

\section{Summing differentiable paths}\label{sec:Differentiable paths}

%Apparently the definition of a theory with differentiable paths {\it{only}}
%has to be made explicit within some kind of quantitative model. 
We will use maximal simplicity and thus work in $1+1$ dimensions, and will shift from real to imaginary time and vice versa when adequate. 

%%%%%%%%%%%%%%%%%%%%%%%%%%%%%%%
\subsection{Defining a model for differentiable paths} \label{subsec:Defining a model for differentiable paths}
%%%%%%%%%%%%%%%%%%%%%%%%%%%%%%%
Consider the propagation of a non-relativistic particle from some initial position $x_1$ at time $t=0$ to some final position $x_2$ at $t=T$, where $T$ is assumed to be a macroscopic time scale.
As usually done in \cite{FeyHib1965} an arbitrary path $y(t)$ connecting these points 
may be decomposed as $\bar{x}(t)+x(t)$, where $\bar{x}(t)$ is the classical trajectory from $x_1$ to $x_2$
(i.e. $\bar{x}(0)=x_1$ and $\bar{x}(T)=x_2$)
and $x(t)$ is the deviation from that fixed classical path (so $x(0)=x(T)=0$).
Then, if the action is at most quadratic in the position and velocity, $S[x+\bar{x}]=S[x]+S[\bar{x}]$ and thus the path integral factorizes as the product of the exponential of the classical action $S[\bar{x}]$ times a genuine path integral with null boundary conditions which depends only on $T$ but not on the external points, and the Kernel of going from $(x_1,0)$ to $(x_2,T)$ is
\begin{align}\label{PI0}
K(x_1,0;x_2,T)%={\mathcal{N}}\int_{y(0)=x_1}^{y(T)=x_2} \mathcal{D}y(t) e^{i S[y]/\hbar}
={\mathcal{N}}e^{iS[\bar{x}]/\hbar}\int_{x(0)=0}^{x(T)=0} \mathcal{D}x(t) e^{iS[x]/\hbar}
 \quad,
\end{align}
where ${\mathcal{N}}$ is a normalization factor that will be addressed in section \ref{Free Particle: fixing the normalization}. 
Please note that the issue of restricting the deviation $x(t)$ rather than the ``absolute" positions $\bar{x}(t)+x(t)$ is a subtle one\footnote{ Consider a PI in which we first factorize the classical action $S[\bar{x}]$, and then impose some restriction $\mathcal{R}[x]$ on the deviation $x(t)$ from the classical path. This will correspond to the same PI over the total paths $x+\bar{x}$ but with a \textit{different} restriction, say $\mathcal{R}'[x+\bar{x}]$
\begin{align}\label{}\nonumber
 e^{iS[\bar{x}]}\int_{\mathcal{R}[x]} \mathcal{D}x\ e^{iS[x]} = \int_{\mathcal{R'}[\bar{x}+x]} \mathcal{D}(\bar{x}+x)\ e^{iS[\bar{x}+x]}
\end{align}
and thus ``simple" restriction $\mathcal{R}[x]$ (such as that presented in this paper) may translate as a very complicate restriction $\mathcal{R'}[x+\bar{x}]$ over the original space. %In Feynman's approach this topic is of no importance, since one sums over all functions, whereby a change of variable has no effect
}.
Based on (\ref{PI0}), path integrals may be constructed by integrating over the particle's intermediate positions (configuration space) or by integrating over the Fourier coefficients that define the paths. It turns out that a differentiability condition is most naturally expressed as a restriction over the Fourier amplitudes, and since $x(t)$ has null borders, it is convenient to express the deviation from the classical trajectory as a Fourier series
\begin{align}\label{xvonT}
x(t)=\sum_{n=1}^\infty a_n \sin\left( \frac{n\pi t}{T} \right) \quad.
\end{align}
In this paper, we will focus on  controlling only the deviation $x(t)$ from the classical path: 
the classical path $\bar{x}$ (which depends on the end-points)
is not modified in any way, but rather restrictions are only imposed over the ``quantum" deviations $x(t)$ (which are end-points-independent). For this reason, the net effect of this restriction over the propagator will be to transform the Feynman propagator into a ``differentiable-path integral" (DPI) propagator (hereafter called D-propagator) according to
\begin{align}\label{KD=KFPi}
K_D(x_1,0;x_2,T)=K_F(x_1,0;x_2,T)\cdot \Pi(T)
\end{align}
where $\Pi(T)$ is to be computed for each potential, and which depends only on time (and possibly other parameters $A$ and $\alpha$ introduced below) but not on the end points. 

The central theorem about Fourier series which we will rely on is the following \cite{CH1924,Jack1920}, which relates the degree of differentiability of a function with the decay of its Fourier coefficients:
\begin{itemize}
\item Theorem:  Let $x(t)=\sum_n a_n e^{int}$ be represented as a Fourier series, and $\alpha\in \mathbb{N}_0 $. Then 
\begin{align}\label{Restriction}
| a_n| \leq \frac{A}{|n|^{\alpha}}\ \ \mbox{for some }\ A>0\ \ \Longleftrightarrow\ \ x\in C^{\alpha-1}
\end{align}
 where $C^\alpha$ is the space of functions with $\alpha$ continuos derivatives. Here $A$ is independent of ``momentum" $n$ and has dimension of length.
\end{itemize}
In principle, one would desire a method by which to include all $C^\alpha$ functions into the path integral (and thus possibly avoiding the introduction of the constant $A$) and thus having a closed vector space. However the authors have found no way of doing that. Therefore we shall proceed in a much more narrow way, by imposing condition \eqref{Restriction} over the path integral. In other words, this restriction amounts to summing over a very narrow subset of all $C^{\alpha}$ functions.
%\footnote{The difficulty is that any arbitrarily large but finite $A$ can enter \eqref{Restriction}, and only $A=\infty$ (which is the Feynman case) spoils the differentiability. Actually, one might be not really interested in making the paths  smooth, but one might 
%wish to make the action finite.  }
Controlling $\alpha$ appropriately will enable us to control the nature of paths entering the PI.  In this restricted model $A$ and $\alpha$ are the two free and continuos control parameters. \\

It is instructive to classify the kinds of paths for finite $A$ and discrete values of $\alpha$ (for $0<\eta<1$)
\be\label{identification}
x(t)\quad {\mbox{subject to (\ref{Restriction}) for}}\quad
\left\{
\begin{array}{cc}
\alpha=1:& {C^0\ -\ \mbox{ continuos \ \ \ \ \ \ \ \ \ \ \ \ }}\\
\alpha=2:& {C^1\ -\ \mbox{ once differentiable }}\\
\alpha=3:& {C^2\ -\ \mbox{ twice differentiable }}\\
\vdots
\end{array}
\right.
\ee

%The two control parameters $A$ and $\alpha$ posses a qualitative difference: the sum in \eqref{alphak} will always diverge for any $\alpha$ as $A\rightarrow \infty$, but for finite $A$ diverges only when $\alpha <1$ (a finite value ``critical point").\\

In practice, the effect of restricting the paths will modify the usual (Feynman) measure:
\be\label{FeynFour}
 \int_F \mathcal{D} x(t) \sim \prod_n \int_{-\infty}^{\infty} dx_n \sim \prod_n \int_{-\infty}^{\infty} da_n\quad,
\ee
which will become, due to restriction \eqref{Restriction}, a D-measure
\be\label{DiffFour}
\int_D \mathcal{D} x(t) \sim \prod_n \int_{-A/n^{\alpha}}^{A/n^{\alpha}} da_n \quad,
\ee

Thus $A$ is a key control parameter 
that allows to return to the standard PI formulation at any stage by taking $A\rightarrow \infty$. Nevertheless $A$ need not necessarily to be a constant: in general it may be a function of the other physical parameters, and particularly relevant is to keep in mind its possible time dependence $A=A(T)$ meaning that the space of allowed paths may also evolve dynamically. We will show one way to fix the function $A(T)$ in \ref{supar:DiffScale}.

%%%%%%%%%%%%%%%%%%%%%%%%%%%%%%%55
\subsection{Upper bounds on physical quantities}
\label{Glimpse}
%%%%%%%%%%%%%%%%%%%%%%%%%%%%%%%55
We now describe briefly the effect of the differentiability exponent $\alpha$ on the ``kinematics"\ of a quantum theory. \\

Let's illustrate \eqref{identification} for the quantum deviation (\ref{xvonT}) from the classical trajectory
subject to the restriction 
\begin{align}\label{restr}
|a_n|\leq \frac{A}{n^{\alpha}}
\end{align}

First, the absolute value of this deviation $|x(t)|$ is bounded at any time by 
\begin{align}\label{x(t)}
|x(t)|%=\Big| \sum_{n=1}^\infty a_n \sin\left( \frac{n\pi t}{T} \right) \Big| \leq  \sum_{n=1}^\infty |a_n|
\leq \sum_{n=1}^\infty \frac{A}{n^{\alpha}} 
= \left\{
     \begin{array}{lr}
       \infty & : \alpha \leq 1 \\
       A\ \zeta(\alpha) & : 1< \alpha
     \end{array}
   \right. \quad ,
\end{align}
where $\zeta(\cdot)$ is the Riemann zeta function
%\footnote{ All along this paper we interpret divergent series as problematic and ill-defined; we make no attempt to ``cure" these infinities by assigning them finite values through their analytic continuation (a method widely used in QFT). }.
The series diverges for $\alpha \leq 1$, indicating the presence of non-continuos (not bounded) paths.
Analogously for the absolute value of the quantum velocity:
\begin{align}\label{vbound}
|\dot{x}(t)|%=\Big| \sum_{n=1}^\infty a_n \frac{n\pi}{T} \cos\left( \frac{n\pi t}{T} \right) \Big| \leq  \sum_{n=1}^\infty \frac{n\pi}{T} |a_n|
\leq \frac{\pi A}{T} \sum_{n=1}^\infty \frac{1}{n^{\alpha-1}}=
\left\{
     \begin{array}{lr}
       \infty & : \alpha \leq 2 \\
       \frac{\pi A}{T}\ \zeta(\alpha-1) & : 2<\alpha
     \end{array}
   \right.\quad.
\end{align}

For example, when $1<\alpha\leq 2$, the velocity is not bounded, although the particle's distance to the origin $|x(t)|$ is always finite. This of course corresponds to fractal trajectories, which possess infinite length. Thus $\alpha=2$ corresponds to the critical value of the differentiability exponent $\alpha$ in order to ensure that $\langle v^2 \rangle $ remains finite. Although these properties have been long known \cite{AbbotWise1980}, the emphasis here relies on the their dependence, in Fourier space, upon the values of the control parameters $A$ and $\alpha$.

%%%%

%%%%%%%%%%%%%%%%%%%%%%%%%%%%%%%55
\section{Squared velocity $\langle v^2\rangle$ and commutator $[x,p]$}\label{sec: General results}

In this section we compute the differentiable-generalization of the quantum square velocity $\langle v^2\rangle$ which in turn allows to derive the modified canonical commutator $[x,p]$. These are considered for the case of free motion.%\footnote{In the usual Feynman case ($A\rightarrow\infty$), if one adds a harmonic oscillator interaction $\sim \omega^2x^2$, it can be shown that $\langle v^2\rangle$ is actually independent of $\omega$.}
%In this sense the results from this section can be considered
%``general'' results of the differentiable model (\ref{Restriction}).

%%%%%%%%%%%%%%%%%%%%%%%%%%%%%%%55
\subsection{A divergence in QM: the mean square velocity $\langle v^2\rangle$}
%%%%%%%%%%%%%%%%%%%%%%%%%%%%%%%55

The arguments pointed out in section \ref{Introduction} were first deduced in \cite{FeyHib1965}, where it is concluded that the quantum mechanical mean square velocity diverges when the time slicing goes to zero $\epsilon \rightarrow 0$ 
\begin{align}\label{v^2FeyHib}
\langle v^2 \rangle =\langle \left( \frac{x_{k+1}-x_k}{\epsilon} \right)^2 \rangle=-\frac{\hbar}{im\epsilon} \quad.
\end{align}

Of course, this divergence is not ``surprising" in the Feynman method, since $x_{k+1}$ and $x_k$ are independent variables.
This is usually deduced in a somewhat complicated and indirect way (as a by-product of computing transition elements). 
The time slicing $\epsilon$ acts as a ``resolution scale", probing the fractal nature of Feynman's method.\\

In this section an alternative method to derive $\langle v^2 \rangle $ is presented, which is performed straightforwardly in Fourier space, a procedure that will prove to be more suitable when attempting the modification of the underlying theory. To compute the mean square velocity for the free particle, we first select an intermediate time $0\leq t_0\leq T$ and for each path $x(t)$ we sum the square of its velocity at $t_0$ $v(t_0,\epsilon)=\frac{x(t_0+\epsilon)-x(t_0)}{\epsilon}$ weighted by $e^{-\frac{1}{\hbar}S}$, and finally normalize. Given a small but finite $\epsilon$, one calculates (we work in imaginary time here)
\begin{align}\label{}
\langle v^2 \rangle (t_0,\epsilon)=\frac{\int \mathcal{D}x(t)\ e^{-\frac{1}{\hbar}S[x(t)]}v^2(t_0,\epsilon)}{\int \mathcal{D}x(t)\ e^{-\frac{1}{\hbar}S[x(t)]}}\quad,
\end{align}

Now shifting to Fourier space, the free euclidean action for $x(t)=\sum_n a_n \sin\left( \lambda_n t \right)$ with $\lambda_n=\left( \frac{n\pi}{T} \right)^2$ is
\begin{align}\label{freeS}
S[x]=%\frac{m}{2}\int_0^T dt\ \dot x^2=
\frac{m}{2}\int_0^T dt\ x(t)\left( -\partial_t^2 \right)x(t)=\frac{mT}{4}\sum_n^\infty \lambda_n a_n^2\quad.
\end{align}
since the boundary term vanishes. The velocity is the limit of
\begin{align}\label{}
v(t_0,\epsilon)=\frac{x(t_0+\epsilon)-x(t_0)}{\epsilon}=\frac{1}{\epsilon} \sum_n a_n \left( \sin\left( \frac{n\pi(t_0+\epsilon)}{T} \right)-\sin\left( \frac{n\pi t_0}{T} \right) 
\right)\quad.
\end{align}
therefore the mean value to compute is
\begin{align}\label{31}
\langle v^2 \rangle (t_0,\epsilon)=\frac{1}{\epsilon^2}\frac{\int da_1 da_2\hdots e^{-\frac{m T}{4\hbar}\sum_n \lambda_n a_n^2 } \left( \sum_j a_j   \Delta s_j \right)^2 }{\int da_1 da_2\hdots e^{-\frac{m T}{4\hbar}\sum_n \lambda_n a_n^2 }  }
\end{align}
where
\begin{align}\label{Dsj}
\Delta s_j=\sin\left( \frac{j\pi(t_0+\epsilon)}{T} \right)-\sin\left( \frac{j\pi t_0}{T} \right)\quad.
\end{align}

Now note that the squared sum in \eqref{31} contains even and odd terms, $\left( \sum_j a_j \Delta s_j \right)^2=(a_1 s_1)^2+\hdots + 2\left[ a_1a_2+\hdots \right]$
and all the odd terms will vanish upon integration, leaving only quadratic terms in the numerator. Thus in \eqref{31} all factors in the numerator cancel with their their equivalents in the denominator, except for the one selected in the sum, yielding 
%\begin{align}\label{}
%\langle v^2 \rangle (t_0,\epsilon)=\frac{1}{\epsilon^2}\frac{\int da_1 da_2\hdots e^{-\frac{m T}{4\hbar}\sum_n \lambda_n a_n^2 } \sum_j (a_j   \Delta s_j )^2 }{\int da_1 da_2\hdots e^{-\frac{m T}{4\hbar}\sum_n \lambda_n a_n^2 }  }
%\end{align}

%The numerator is:
%\begin{align}\label{}
%&=\sum_j \int da_1 e^{- \frac{m T}{4\hbar} \lambda_1 a_1^2}\int da_2 e^{- \frac{m T}{4\hbar} \lambda_2 a_2^2}\hdots \left( \int da_j e^{- \frac{mT}{4\hbar} \lambda_j a_j^2} a_j^2 \Delta s_j^2 \right)\hdots 
%\end{align}
%so once divided by the denominator all factors cancel out, except the one selected in the sum, yielding
\begin{align}\label{v2general}
\langle v^2 \rangle (t_0,\epsilon)= \frac{1}{\epsilon^2}\sum_j \Delta s_j^2 \frac{\int da_j\ a_j^2\ e^{- \frac{mT}{4\hbar} \lambda_j a_j^2}  }{\int da_j\ e^{- \frac{m T}{4\hbar} \lambda_j a_j^2} }\quad.
\end{align}

%
%\begin{align}\label{}
%\langle v^2 \rangle (\epsilon)= \frac{1}{\epsilon^2}\sum_{j=1}^\infty \sin^2 \left( \frac{j\pi\epsilon}{T} \right) \frac{\int da_j e^{- \frac{m}{2\hbar} \frac{T}{2}\lambda_j a_j^2} a_j^2 }{\int da_j e^{- \frac{m T}{4\hbar} \lambda_j a_j^2} }
%\end{align}
This is the series we will evaluate in the next subsections, for both the Feynman case and for the differentiable version of it. 

%%%%%%%%%%%%%%%%%%%%%%%%%
\subsubsection{The Feynman case}
Having no restriction on the integrals in \eqref{v2general}, simple gaussian integration yields
\begin{align}\label{}
\frac{\int_{-\infty}^{\infty} da_j\ a_j^2\ e^{- \frac{m T}{4\hbar} \lambda_j a_j^2}  }{\int_{-\infty}^{\infty} da_j e^{- \frac{m T}{4\hbar} \lambda_j a_j^2} }%=\frac{1}{2 \frac{mT}{4\hbar}  \lambda_j}
=\frac{2\hbar}{mT} \left( \frac{T}{j \pi} \right)^2  \quad.
\end{align}
so plugging this into \eqref{v2general} and replacing $\Delta s_j$ from \eqref{Dsj}, we get
\begin{align}\label{v2Fseries}
\langle v^2 \rangle _F(\epsilon)&=\frac{2\hbar}{mT}\left( \frac{T}{\pi\epsilon} \right)^2\sum_{j=1}^\infty \frac{1}{j^2}\left[ \sin\left( \frac{j\pi(t_0+\epsilon)}{T} \right)-\sin\left( \frac{j\pi t_0}{T} \right) \right]^2
\end{align}
This series as a function of $\epsilon$ must be treated with care. In Appendix \ref{Feynman Series} it is shown to be independent of $t_0$, and its first order Taylor expansion around $\frac{\epsilon}{T}=0$ is derived, and gives
\begin{align}\label{v^2Fey}
\langle v^2 \rangle _F(\epsilon)&=\frac{2\hbar}{mT}\left( \frac{T}{\pi\epsilon} \right)^2\frac{\pi^2}{2}\frac{\epsilon}{T} =\frac{\hbar}{m\epsilon}\quad.
\end{align}
This is precisely the result obtained in \cite{FeyHib1965} in a rather indirect way. Note here that the final result turns out to be independent of both $t_0$ and $T$. 

%%%%%%%%%%%%%%%%%%%%
\subsubsection{Differentiable case} \label{subsec: FreeDifferentiable case}
In this model \eqref{DiffFour}, the analytic calculation becomes much more arduous, for the bounds over the gaussian integrals drop extra factors which are not easy to manipulate. The quotient in \eqref{v2general} gives in this case the same Feynman pre-factor but adds a modification: 
\begin{align}\label{intquotD}
\frac{\int_{-A/j^{\alpha}}^{A/j^{\alpha}} da_j\ a_j^2\ e^{- \frac{m T}{4\hbar}   \lambda_j a_j^2}  }{\int_{-A/j^{\alpha}}^{A/j^{\alpha}} da_j\ e^{- \frac{m T}{4\hbar} \lambda_j a_j^2} }&\equiv \frac{2\hbar}{mT} \left( \frac{T}{j \pi} \right)^2 \cdot \Big( 1- \underbrace{Z(W_j)}_{\mbox{modification}}\Big) 
\end{align}
where $Z$ and $W$ are the dimensionless functions
\begin{align}\label{Z y W QM}
Z(W)=\frac{2}{\sqrt{\pi}} \frac{\sqrt{W} e^{-W} }{\mbox{Erf}(\sqrt{W})}\hspace{2cm}W_j=\sqrt{\frac{m T}{4\hbar}  }\left( \frac{j\pi}{T} \right) \frac{A}{j^{\alpha}}=\frac{\bar{A}}{j^{\alpha-1}}
\end{align}
and $\bar{A}$ is also a dimensionless quantity:
\begin{align}\label{barA}
\bar{A}=\sqrt{\frac{m \pi^2}{4\hbar T}  } A
\end{align}

\begin{figure}[hbt]
   \centering
%\centerline{\protect\vbox{\epsfig{file=Approx.eps,
%width=0.6\textwidth}}}
\includegraphics[width=10cm]{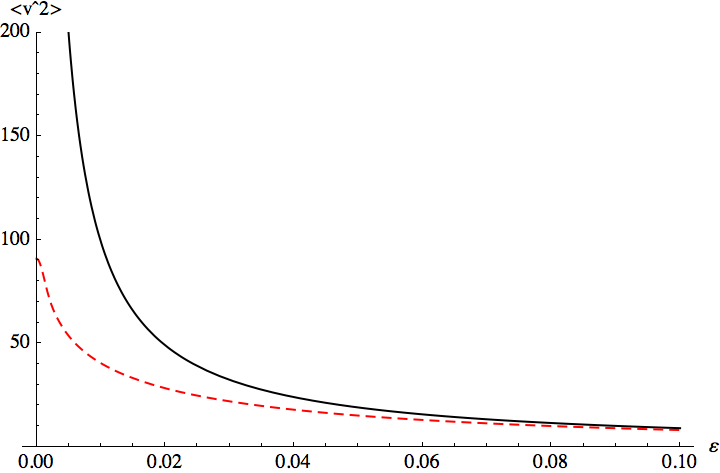}
  \caption{Numerical plot of the Feynman $\langle v^2 \rangle _{F}$ (solid black), and the differentiable $\langle v^2 \rangle _D$ (dashed red) for $\alpha=2.1,A=10,T=1,m=1,\hbar=1$ (in natural units). Note the bifurcation occurs loosely around $\sim 0.03$ }
\label{fig:v2}
\end{figure}

To simplify a bit the calculations, we consider the mean velocity at the origin so set $t_0=0$ (recall that in Feynman's case the result is independent of $t_0$), with which \eqref{Dsj} becomes $\Delta s_j=\sin\left( \frac{j\pi\epsilon}{T} \right)$, so replacing \eqref{intquotD} into \eqref{v2general} yields
\begin{align}\label{55}
\langle v^2 \rangle _D(\epsilon)
&=\frac{2\hbar}{mT} \left( \frac{T}{\pi\epsilon} \right)^2\sum_j  \frac{\sin^2 \left( \frac{j\pi\epsilon}{T} \right)}{j^2} \left( 1-Z(W_j) \right)
\end{align}
so we must find this series as a function of the time interval $\epsilon$. But before showing to the analytic results, we show in Figure \ref{fig:v2} a plot comparing numerically the usual UV-divergent $\langle v^2 \rangle _F$ \eqref{v^2Fey} with the modified version $\langle v^2 \rangle _D$ \eqref{55}.
%This plot illustrates that, for any fixed value of the external parameters $\alpha,A,T,m$, there exists a special value of the time scale, say $\epsilon_D$ (in the plot, around $0.03$) above which the differentiable curve is very close to the Feynman one, and below which they are very different, as Feynman's velocity is divergent while the differentiable one is convergent. We shall call $\epsilon_D$ (or more properly, the quotient $\epsilon_D/T$) the  differentiable scale: if the experimental time interval $\epsilon$ with which we are probing nature is ``coarser", $\epsilon_D<\epsilon$ (low resolution) we find (approximately) Feynman's curve an thus usual QM; if however we perform experiments for very short times, $\epsilon<\epsilon_D$ (high resolution) we find a different theory, one which is UV-convergent. \\

An analytic estimation of \eqref{55} is done extensively in Appendix \ref{Differentiable Series}, arriving at the following result (for $\alpha>2$)
\begin{align}\label{v2D}
   \langle v^2 \rangle _D = \left\{
     \begin{array}{lr}
     v^2_{UV}  \hspace{1.5cm} \epsilon \ll \epsilon_D \hspace{1.5cm}v^2_{UV}=\frac{\pi A}{T} \sqrt{\frac{\hbar}{m T}} \\
      \\
       \frac{\hbar}{m\epsilon}-\frac{C}{\epsilon^2}  \hspace{1cm} \epsilon_D<\epsilon \hspace{1.5cm}C=\frac{4}{\pi^3}\frac{1}{A}\left( \frac{\hbar T}{m} \right)^{3/2} 
     \end{array}
   \right. 
\end{align} 
which  regularizes the quantum velocity in the UV, converging to a finite value $v_{UV}$. Next we give a geometrical interpretation of the differentiable scale $\epsilon_D$, and show how to compute it as function of the other parameters of the theory.

\subparagraph{The differentiable time scale $\epsilon_D$}\label{supar:DiffScale}  \ \\

To fully understand the emergence of a microscopic differentiable scale, it is useful to recall the basic property of nowhere-differentiable/fractal functions: they obey certain self-similarity scaling laws  at all scales, no matter how small. For the case of usual QM, that is $\langle v^2 \rangle \sim \frac{1}{\epsilon}$ or $\langle \Delta x \rangle  \sim \sqrt{\epsilon}$ as a Brownian motion. In the differentiable model these scaling laws are valid only above a given time scale, $\epsilon_D$, making sample paths to  appear as being fractals, but change for times shorter than $\epsilon_D$, revealing the differentiable properties of paths at those scales.\\

A simple example will serve to illustrate how this works. Consider a ``Feynman" quantum sample path from $(0,0)$ to $(0,T)$, which is known to behave as a Brownian walk, written as a random Fourier series \cite{Wiener1923,CuLe1980}
\begin{align}\label{Brownian}
X_F(\tau)= \sqrt{\frac{\hbar T}{m}} \sum_{j=1}^\infty \frac{N_j}{j}\sin( j\tau )%\hspace{1cm}\tau=\frac{\pi\epsilon}{T}
\end{align}
where $N_j$ are independent equally distributed real numbers with mean zero, and the pre factor $\sqrt{\frac{\hbar T}{m}}$ gives the length dimension. 
%\footnote{This is not purely by dimensional analysis. Actually one can compute, with the exact same method illustrated in this section, the mean square distance, and finds:
%\begin{align}\label{note}
%\langle x^2(t)\rangle=\frac{\hbar T}{m}\sum_{j=1}^\infty \sin^2\left( \frac{j\pi t}{T} \right) \frac{2}{j^2\pi^2}
%\end{align}}
For simplicity, assume $|N_j|\leq 1$ for this example. It is well known that \eqref{Brownian} converges uniformly to a continuos but nowhere-differentiable function of fractal (Hausdorff) dimension $d=3/2$. Now let's define a ``differentiable version" called $X_D(\tau)$: a function that looks very ``similar"\ to $X_F(\tau)$ when examined with low time resolution $\delta \tau \geq \tau_D$, therefore  seeming to be nowhere-differentiable (i.e. obeying the same scaling laws), but whose differentiable nature becomes evident as we probe it with higher resolutions $\delta \tau\leq \tau_D$.\\

\begin{figure}[hbt]
   \centering
%\centerline{\protect\vbox{\epsfig{file=Approx.eps,
%width=0.6\textwidth}}}
\includegraphics[width=10cm]{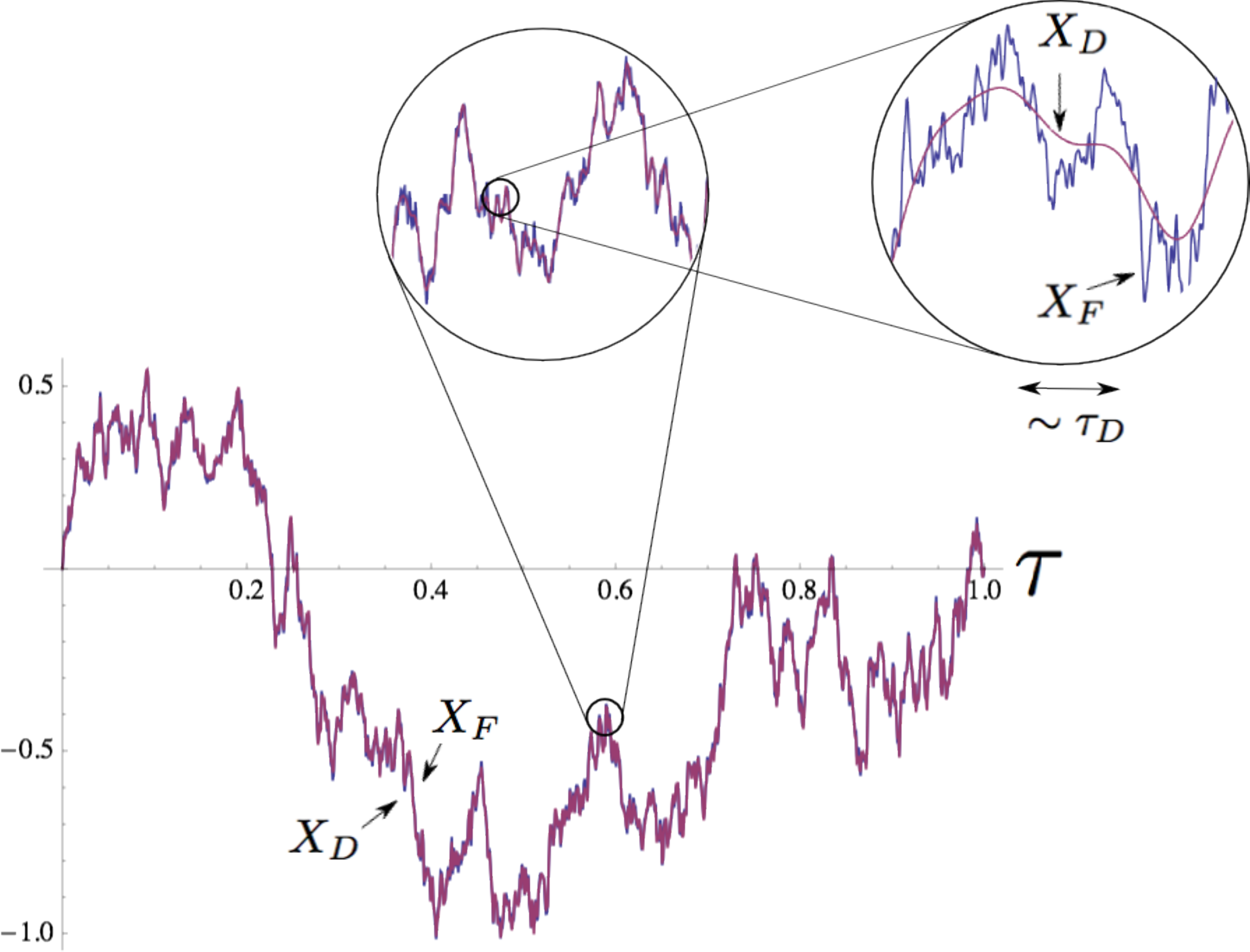}
  \caption{Illustration of the scale dependence of ``fractality": a path which is actually differentiable may appear as fractal at coarser scales, but eventually one reaches the scale $\tau_D$ where its differentiable properties become manifest. }
\label{fig:Brown}
\end{figure}

 To construct the differentiable function $X_D(\tau)=\sum a_j \sin(j \tau)$, note that for a given value of $A$, a certain amount of the first coefficients $a_j$, say $j=1,\dots, j_D$, may be taken as identical to those of $X_F(\tau)$,  for as long as the differentiable restriction \eqref{restr} is satisfied, that is:
 \begin{align}\label{jD}
\sqrt{\frac{\hbar T}{m}}\frac{1}{j}\leq \frac{A}{j^\alpha}\hspace{.5cm}\mbox{for}\hspace{.5cm}j=1,\dots,j_D\hspace{1cm}\Rightarrow \hspace{1cm} j_D^{\alpha-1}=\frac{A}{\sqrt{\frac{\hbar T}{m}}}
\end{align}
so the path is defined as:
\begin{align}\label{}
X_D(\tau)=\sum_{j=1}^\infty a_j \sin(j \tau)\hspace{1cm}\mbox{with}\ \ a_j= \left\{
     \begin{array}{lr}
       \sqrt{\frac{\hbar T}{m}}\cdot \frac{N_j}{j}\hspace{.5cm} \leq \frac{A}{j^{\alpha}} & : j \leq j_D \\
       a_j \hspace{1.45cm} \leq \frac{A}{j^{\alpha}} & : j_D<j
     \end{array}
   \right.
\end{align}
Figure \ref{fig:Brown} illustrates what's going on geometrically: if one plots both $X_F(\tau)$ and $X_D(\tau)$ together, they are indistinguishable when probed at low resolution/large scales; on must zoom in until a time scale of order $\tau_D$ to start noticing the different nature of both paths. 
Now $j_D$ is the highest frequency at which $X_D$ will ``seem" fractal, so the time scale $\tau_D=\frac{\epsilon_D}{T}$ at which this occurs is its conjugate, so replacing $j_D=\tau_D^{-1}$ in \eqref{jD} one finds the differentiable scale $\epsilon_D$ as function of the parameter $A$:
\begin{align}\label{epsilonA}
\left( \frac{T}{\epsilon_D} \right)^{\alpha-1}=\frac{A}{\sqrt{\frac{\hbar T}{m}}}
\end{align}

Equation \eqref{epsilonA} relates two external unknown quantities to be determined, $A$ and $\epsilon_D$. Of these two, $\epsilon_D$ is more directly connected to a physical experiment: it is the time interval at which one expects to find new physics, while $A$ is a less intuitive quantity related to the allowed space of paths. If we choose to fix $A=A(T)$ in an arbitrary manner, then in general \eqref{epsilonA} would give a time dependent $\epsilon_D=\epsilon_D(T)$ which would imply that $\epsilon_D$ increased or decreased arbitrarily in time: the scale of new physics would change in time, a somewhat bizarre scenario. The most natural 
option is to  impose that $\epsilon_D$ is a constant %\footnote{Or at least approximately a constant. One must remain open to consider a slight cosmological variation of any of nature's constants.}
whose value must be bounded experimentally, and this fixes the function $A$ (up to $\epsilon_D$)
\begin{align}\label{epsilonA(T)}
A(T)=\sqrt{\frac{\hbar T}{m}} \left( \frac{T}{\epsilon_D} \right)^{\alpha-1}
\end{align}

In \eqref{epsilonA(T)}, if we think of $\epsilon_D$ as the independent parameter, $A(T)$ evolves as the product of $\sqrt{\hbar T/m}$ %(the same in \ref{note}) 
which gives it the length dimension, times the dimensionless amplification factor $\left( T/\epsilon_D \right)^{\alpha-1}$. As expected, $A\rightarrow \infty$ as $\epsilon_D\rightarrow 0$ recovering Feynman's case. Anyway, one must bear in mind that the choice \eqref{epsilonA(T)} is not mandatory, as one could follow another criterion that gives another time dependence. Nonetheless, not any election seems plausible because, apart from the above mentioned arguments, the form $A(T)$ decides the unitarity of the theory. As we show in subsection \ref{subsec:The question of Unitarity}, \eqref{epsilonA(T)} does respect unitarity, while others choices will violate it.

%%%%%%%%%%%%%%%%%%%%%%%%%%%%%%%55
\subsection{Uncertainty and Commutation relations: Path Integral point of view}
%%%%%%%%%%%%%%%%%%%%%%%%%%%%%%%55
The study of the commutation and uncertainty relations within the context of path integrals is a topic seldom addressed in the literature. When mentioned, it is commonly stated that the very irregular nowhere-differentiable, fractal nature of the functions entering the path integral are a necessary requisite for reproducing the canonical commutation relations $[x,p]=i\hbar$. In this section we dispute this notion. We argue that this is true only if one wishes that this commutation law remains valid  at all scales, including arbitrarily high energies (short times). Intuitively the results are analogous to those from section \ref{sec: General results}: by summing only over well-behaved controlled paths, one obtains a scale-dependent commutator, which reduces to the usual one for coarse resolution measurements (above the differentiable scale $\tau_D$, where paths ``look like" fractals), but vanishes for time intervals shorter the differentiable time scale.\\

It is the intimate connection between $\langle v^2 \rangle $ and the canonical commutator $[x,p]$ which makes the conclusions of the previous section also applicable here. This link was provided in \cite{FeyHib1965}: given $\langle v^2 \rangle =\langle \big( \frac{(x_{k+1}-x_k)}{\epsilon} \big)^2\rangle$ we can derive the commutation relations in the following way 
\begin{align}\label{}
\epsilon\langle v^2 \rangle &=\langle    \frac{(x_{k+1}-x_k)x_{k+1}}{\epsilon} \rangle -\langle    \frac{(x_{k+1}-x_k)x_k}{\epsilon} \rangle 
\end{align}
but the first term may be approximated by $\langle  (x_{k+1}-x_k)x_{k+1}\rangle \approx \langle  (x_k-x_{k-1})x_k\rangle +\mathcal{O}(\epsilon)$ since it is exactly the same quantity but evaluated at a time $\epsilon$ before, yielding 
\begin{align*}
%\epsilon\langle v^2 \rangle &= \langle   x_{k} \frac{(x_{k}-x_{k-1})}{\epsilon} \rangle -\langle    \frac{(x_{k+1}-x_k)}{\epsilon} x_k\rangle \\
m\epsilon\langle v^2 \rangle &=\langle  x_k p_{k-1}\rangle -\langle  p_k x_k \rangle ``=" \langle [x,p]\rangle 
\end{align*}
where we have identified quantities evaluated at $k-1$ as being  previous than $k$, and thus operating  first (to the right). Therefore one concludes that
\begin{align}\label{v^2[x,p]}
m\epsilon\langle v^2 \rangle =\langle [x,p]\rangle
\end{align}

%In the differentiable model this topic is of great importance as it builds a bridge between the path integral and canonical approaches, and therefore translates the differentiability modification (purely path-like motivated) into the canonical language. Recall that the uncertainty relations and the commutator are closely related through
%\begin{align}\label{}
%\Delta x\Delta p \geq \frac{1}{2}\big|\langle [x,p] \rangle\big|
%\end{align}
%where $\langle \cdot\rangle$ here denote a wave function expected value (not a path integral mean value), which in this context is irrelevant. 
 
\subsubsection{Feynman case}

From \eqref{v^2Fey} we found that $\langle v^2 \rangle _F=\frac{\hbar}{m\epsilon}$ by summing over  all paths, yielding through \eqref{v^2[x,p]} the usual commutator and uncertainty:
\begin{align}\label{}
\langle [x,p]\rangle_F=\hbar\hspace{.5cm}\Rightarrow\hspace{.5cm}\left( \Delta x \Delta p\right)_F\geq \frac{1}{2}\hbar
\end{align}
%(an $i$ is missing because we are in imaginary time).

\subsubsection{Differentiable case}\label{v2CCRDifferentiable case}

When we restricted to summing paths only over subclasses of differentiability, we found an approximation for $\langle v^2 \rangle _D$ in \eqref{v2D}, which is comprised of two distinct behaviors for two different regimes. Using \eqref{v^2[x,p]} we see that the modified uncertainty-commutator is:
\begin{align}\label{dxdpD}
   \langle [x,p]\rangle_D = \left\{
     \begin{array}{lr}
     m\epsilon\ v^2_{UV} & : \epsilon \ll \epsilon_D   \\
      \\
       \hbar-\frac{mC}{\epsilon} & : \epsilon_D<\epsilon \\
     \end{array}
   \right. 
\end{align} 

This is one of the main results of this paper. For ``long" times/low resolutions $\epsilon\gg  \epsilon_D$, the $\frac{\hbar}{2}$ dominates over the correction, and we recover the usual quantum mechanical relations (with a small modification), and it is in this regime where paths appear to be nowhere-differentiable. However for time scales shorter than the differentiable scale $\epsilon_D$, the uncertainty and commutator vanish as $\sim\epsilon\rightarrow 0$, because the velocity reaches an UV-convergent constant value $v_{UV}$. The interpretation of this latter region is not clear, as it is neither fully ``quantum", nor fully ``classical" for there are still infinite paths contributing to the path integral. \\

In \eqref{dxdpD}, it is not obvious at first sight how to interpret the $\sim \frac{1}{\epsilon}$ correction at low resolution, $\epsilon_D<\epsilon$. However an important insight is gained if we express time $\epsilon$ in terms of momentum, for in this region at first order \eqref{v2D} we have Feynman's velocity \eqref{v^2Fey}, which can be restated in terms of the momentum
\begin{align}\label{ep}
\langle v^2 \rangle _D\approx \frac{\hbar}{m\epsilon}\hspace{1cm}\Rightarrow\hspace{1cm}\frac{m}{\epsilon}\approx  \frac{m^2\langle v^2 \rangle _D}{\hbar}=\frac{\langle p^2\rangle_D}{\hbar}
\end{align}
where we have identified $\langle p^2\rangle =\langle (mv)^2\rangle$. This, inserted into \eqref{dxdpD} and using \eqref{v2D} yields
\begin{align}\label{[x,p]Dlow}
\langle [x,p]\rangle_D=\hbar-\frac{C}{\hbar}\langle p^2\rangle_D
\end{align}
with
%Now notice that $C$ can be related to $v^2_{UV}$ by replacing $A$ from \eqref{v2D} yielding
\begin{align}\label{}
C=\frac{4}{\pi^3}\frac{1}{A}\left( \frac{\hbar T}{m} \right)^{3/2}=\left( \frac{2}{\pi} \right)^2  \frac{\hbar^2}{p^2_{UV}}
\end{align}
where we have defined $p_{UV}=m\ v_{UV}$, the ultraviolet limit of the particle's momentum when measured at infinite resolution $\epsilon\rightarrow 0$ in \eqref{v2D}. Finally, reading \eqref{[x,p]Dlow} into the canonical language (i.e. without the brackets), we have
\begin{align}\label{DGUP}
 [x,p]_D=\hbar \left( 1-\left( \frac{2}{\pi} \right)^2  \frac{p^2}{p^2_{UV}} \right)\hspace{1cm},\ \ \ \mbox{valid for}\ \ \ p<p_D
\end{align}
where $p^2_D\equiv \frac{\hbar m}{\epsilon_D}$, which is the momentum scale associated to the differentiable scale $\epsilon_D$ through \eqref{ep}.\\
 
Whenever curves display fractal behavior, the UV limit of the momentum is infinite (paths are nowhere-differentiable) $p_{UV}=\infty$ and the correction vanishes. It is interesting to note that the coefficient accompanying $p^2$, act's as a  coupling: it is the system's effective low energy ``memory" of its high-energy (short-time) properties. As we shall mention in \ref{sec:Relation to other approaches}, this may have some relation with the concept of Generalized Uncertainty Principle \cite{KMM1995,QuesneTkachuk2007,CLMT2011,DasPramanik2012} which proposes a modified commutator of the form \eqref{DGUP}. %As mentioned in these later papers commutator \eqref{} leads to an uncertainty... {\color{red} Don't know wether to risk or not putting a plot which is not fully understood...}

%%%%%%%%%%%%%%%%%%%%%%%%%%%%%%%55
\section{The modified Schr\"odinger equation: shift of energy levels}\label{sec:SED}
%%%%%%%%%%%%%%%%%%%%%%%%%%%%%%%55
Given its importance in QM, we must derive the D-version of the Schrodinger equation.
As noted in \cite{FeyHib1965}, in the PI formulation this is achieved by first defining the wave function's evolution $\psi(t_1)\rightarrow \psi(t_2)$ 
due to the kernel (\ref{PI0}) as
\be\label{PropWF}
\psi(t_2,x)= \int_{-\infty}^\infty dy\ K(t_1,y,t_2,x) \psi(t_1,y)\quad.
\ee
and then noting that the Feynman kernel $K_F$ satisfies
the Schr\"odinger %\cite{Blau2013}
equation for the final position and time \cite{FeyHib1965}.

%
%\be\label{SchroedKernel}
%i\hbar \frac{\partial}{\partial t}K_F(0,y;t,x)=\hat H_x\ K_F(0,y;t,x)\quad,
%\ee
%where $\hat H_x=-\frac{\hbar}{2m}\partial_x^2+V(x)$ is the usual Hamiltionian,
%and then take $\partial_t$ to \eqref{PropWF} in combination with \eqref{SchroedKernel}, giving the normal wave equation:
%\be
% i\hbar\ \partial_t \psi(t,x)=\int_{-\infty}^\infty dy \;\hat H_x\ K_F(0,y;t,x)\psi(0,y)=\hat H_x \ \psi(t,x)\quad.
%\ee
%This procedure will prove useful in the modified case below. 

\subsubsection{Differentiable case}\label{subsubsec:SEDifferentiable case}

In the DPI model, the propagation is due to the kernel $K_D$ which must enter \eqref{PropWF}. Now, recall from subsection \ref{subsec:Defining a model for differentiable paths} that in the differentiable case the kernel factorizes as a product of the Feynman kernel times a modification $\Pi(t)$ which is basically dependent on time but independent of the external end points; so if a particle propagates from $(y,0)$ to $(x,t)$
\be\label{KD=KDPi}
K_D(0,y;t,x)=K_F(0,y;t,x)\cdot \Pi(t)\quad,
\ee

This allows to derive the generalization of the
Schr\"odinger equation for the differentiable case, by taking $i\hbar\ \partial_t$ to \eqref{PropWF} with $K=K_D$, replacing \eqref{KD=KDPi} and using that $K_F$ satisfies the usual wave equation, one finds:
\bea
i\hbar\ \partial_t \psi(t,x)&=& i\hbar \int_{-\infty}^\infty dy\; \Big( \partial_t K_F(0,y;t,x)\cdot \Pi(t)
+K_F(0,y;t,x)\cdot  \partial_t \Pi(t)\Big) \psi(0,y)\\ \nonumber
%&=&\int dy \;\left( \hat H_x K_F(0,y;t,x)\cdot \Pi(t,A,...)
%+i\hbar\ K_F(0,y;t,x)\cdot  \partial_t \Pi(t,A,...)\right) \psi(0,y)\\ \nonumber
%&=&  \hat H_x \int_{-\infty}^\infty dy\ K_F(0,y;t,x) \Pi(t) \psi(0,y)
%+i\hbar \frac{\partial_t \Pi(t)}{\Pi(t)}\int_{-\infty}^\infty dy\ K_F(0,y;t,x) \Pi(t) \psi(0,y)\\ \nonumber
&=&\hat H_x\ \psi(t,x)+i\hbar\frac{\partial_t \Pi}{\Pi}\psi(t,x)
\eea
and thus the DPI wave equation is
\begin{align}\label{DSE}
i\hbar\ \partial_t \psi=\hat H \psi+i\hbar\ \psi\ \partial_t \ln \left( \Pi(t) \right)
\end{align}

This equation may be understood in different ways:

\begin{itemize}

\item The natural interpretation of \eqref{DSE} is the appearance of a modified ``differentiable" Hamiltonian, $\hat{H}_D=\hat{H}+i\hbar\ \partial_t \ln \left( \Pi(t) \right)$, which corresponds to a ``differentiable" potential which in general is time-dependent and complex: $V_D=V+i\hbar\ \partial_t \ln \left( \Pi \right)$. But this immediately rises the question of unitarity in this model: for the time evolution operator $\hat{U}_D(t)=e^{i\hat{H}_Dt/\hbar}$ to be unitary, the Hamiltonian must be Hermitian, $\hat{H}_D^\dag=\hat{H}_D$, which implies that the modification $i\hbar\ \partial_t \ln (\Pi)$ must be real, i.e. $\ln(\Pi)$ must be purely imaginary (in the real-time path integral), that is $|\Pi(t)|=1$. The same conclusion is arrived at by analysing the DPI continuity equation. 

\item An alternative point of view is gained rewriting \eqref{DSE}, by defining $\psi(x,t)=\phi(x,t)\Pi(t)$, which yields
\begin{align}\label{}
%i\hbar\ \partial_t \left( \phi\ \Pi \right)&=\left( \hat H+i\hbar \frac{\partial_t \Pi}{\Pi} \right)\phi\ \Pi\\
i\hbar\left( \phi\ \partial_t \Pi+\Pi\ \partial_t \phi \right)&=\hat H \phi \Pi +i\hbar\ \phi\ \partial_t \Pi\\
i\hbar\ \partial_t \phi=\hat H\phi\ \ &\Leftrightarrow\ \ i\hbar\ \partial_t \left( \Pi^{-1}\psi \right)=\hat H \left( \Pi^{-1}\psi \right)\label{PIpsiSE}
\end{align}
thus we see that it is $\phi= \Pi^{-1}\psi $ which satisfies the ordinary Schrodinger equation. Therefore the product $\Pi^{-1}\psi$ possesses all the usual properties which we normally address to the wave function, and which we know how to compute.

\end{itemize}

What occurs to the energy eigenvalues? Conventionally  time-independent energy spectrums  become time-dependent, but the  spacing between levels remains unchanged always. Given a time-independent Hamiltonian $\hat{H}$, we know from \eqref{PIpsiSE} that the product $\phi=\Pi^{-1}\psi$ is the solution to the usual time-independent Schrodinger equation $\hat{H} \phi=E_S\phi$ (S for Schrodinger) and is expressed as
\begin{align}\label{}
\phi(x,t)=u(x)e^{-\frac{i}{\hbar}E_S t}
\end{align}
where $u(x)$ is the solution to the time-independent eigenvalue equation. Thus the  true eigenvalue associated to the wave function $\psi$ is not $E_S$ (the conventional one) but the result of collecting the extra time dependence from $\Pi(t)$ as an exponential: 
\begin{align}\label{}
\psi(x,t)=\phi(x,t)\Pi(t)=u(x)e^{-\frac{i}{\hbar}E_s t} e^{-\frac{it}{\hbar} \frac{i\hbar}{t}\ln(\Pi(t))}
\end{align}
therefore the DPI energy levels are time dependent and given by
\begin{align}\label{E_DSE}
E_D(t)=E_S+\frac{i\hbar}{ t}\ln \left( \Pi(t) \right)
\end{align}
where $E_S$ is the eigenvalue solution to the usual Schrodinger equation, and the second factor is due to the differentiability restriction. The limit into conventional QM is achieved as always by taking $A\rightarrow\infty$ implying $\Pi (t)\rightarrow 1$. This procedure is completely general, and applicable to any model of modifying the Kernel of QM in the form \eqref{KD=KFPi}.

%%%%%%%%%%%%%%%%%%%%
\section{Concerning unitarity}\label{subsec:The question of Unitarity}
%%%%%%%%%%%%%%%%%%%%

As was outlined in subsection \ref{subsubsec:SEDifferentiable case}, in general a method of restricting the space of functions entering the path integral is not guaranteed to preserve unitarity. Here it is briefly shown that the criterion for respecting unitarity, in the context of the presented model, is that the modifying factor $\Pi(T)$ must be unitary (in real time). Recall from the previous sections that in this method the usual Feynman kernel $K_F$ is replaced by the DPI kernel $K_D$ as $K_F\rightarrow K_D=K_F\ \Pi(T)$. Next one examines the effects of the presence of this $\Pi$ factor in two key relations regarding unitarity: conservation of probability and the propagator time decomposition property (also known as Einstein-Smoluchowski-Kolmogorov-Chapman relation). 

\subsection{Conservation of Probability}
This means that for any times $t$ and $T$:
\begin{align}\label{}
\int_{-\infty}^\infty dx\ |\psi(x,t)|^2=\int_{-\infty}^\infty dx\ |\psi(x,T)|^2
\end{align}
which, by direct use of \eqref{PropWF}, requires that 
%\begin{align}\label{}
%\int_{-\infty}^\infty dx\ |\psi(x,T)|^2%&=\int dx \left| \int dy\ \psi(y,t) K(y,t;x,T) \right|^2\\
%&=\int_{-\infty}^\infty dx \int_{-\infty}^\infty dy\ \psi(y,t) K(y,t;x,T) \int_{-\infty}^\infty dy'\ \psi^*(y',t) K^*(y',t;x,T)\\
%&=\int_{-\infty}^\infty dy\ dy'\ \psi(y,t)   \psi^*(y',t) \int_{-\infty}^\infty dx\ K(y,t;x,T)K^*(y',t;x,T)=\int_{-\infty}^\infty dy\ |\psi(y,t)|^2
%\end{align}
%which requires that
\begin{align}\label{}
\int_{-\infty}^\infty dx\ K(y,t;x,T)K^*(y',t;x,T)=\delta(y-y')
\end{align}
a property that is indeed satisfied by the Feynman kernel $K_F$; in the DPI case it is direct to see that 
one can see that
\begin{align}\label{}
\int_{-\infty}^\infty dx\ K_D(y,t;x,T)K_D^*(y',t;x,T)%&=\Pi(T-t)\Pi^*(T-t)\int_{-\infty}^\infty dx\ K_F(y,t;x,T)K_F^*(y',t;x,T)\\
&=|\Pi(T-t)|^2\ \delta(y-y')
\end{align}
and thus it is needed that $\Pi(\cdot)$ be unitary:
\begin{align}\label{PiExpiPhi}
%|\Pi(T-t)|^2=1\ \ \ \Rightarrow\ \ \ \ 
\Pi(T)=e^{i\varphi(T)}\hspace{.5cm},\hspace{.5cm}\varphi(T)\in \mathbb{R}
\end{align}

\subsection{Kernel decomposition (ESKC relation)} 
This relation requires that a propagator can be expressed as the convolution 
\begin{align}\label{KKK}
K(y,t;x,T)=\int_{-\infty}^\infty dy'\ K(y,t;y',t')K(y',t';x,T)
\end{align}
for any intermediate time $t<t'<T$, a property that is satisfied by the Feynman kernel. 
In the DPI case one needs that 
\begin{align}\label{}
\Pi(T-t)\ K_F(y,t;x,T)=  \Pi(t'-t)\Pi(T-t') \int_{-\infty}^\infty dy'\ K_F(y,t;y',t')K_F(y',t';x,T)
\end{align}
that is, $\Pi$ must be an exponential function:
\begin{align}\label{PiExpOT}
%\Pi(T-t)=\Pi(t'-t)\Pi(T-t)\ \ \Rightarrow\ 
\Pi(T)=e^{\tilde{\Omega} T}
\end{align}
where $\tilde{\Omega}$ is in general a complex constant. \\

From \eqref{PiExpiPhi} and \eqref{PiExpOT}, one concludes that the only form for the modification $\Pi(T)$ that respects both probability conservation and the ESKC relation is
\begin{align}\label{eialphat}
\Pi(T)=e^{i\Omega T}
\end{align}
where $\Omega$ is a real frequency. 
%Interestingly enough, in light of the findings from subsection \ref{subsubsec:SEDifferentiable case} (i.e. that the product $\phi(x,t)=\Pi^{-1}\psi(x,t)$ satisfies the usual Schrodinger equation with no modifications) and \eqref{eialphat}, we see that if unitarity is preserved, then insofar as the wave function is concerned, the net result of the DPI model is that  the actual wave function $\psi$ of a system is a $U(1)$ group transformation of the ``usual" wave function $\phi$ (solution to the normal wave equation), 
%\begin{align}\label{unitPhi}
%\Psi(x,t)\rightarrow \psi(x,t)=e^{i\Omega t}\Psi(x,t)
%\end{align}
%where the phase depends only on time and not on space. 
A unitary evolution as \eqref{eialphat} would allow
only for constant shifts in the energy levels of a bound system. 
Due to the fact that a priori we do not know the functional form of $\Pi(t)$, this condition
has to be checked by evaluating explicit examples in section \ref{sec: Specific results}.

%%%%%%%%%%%%%%%%%%%%%%%%%%%%%%%55
\section{The free particle and the harmonic oscillator}\label{sec: Specific results}
In this section the method developed in previous sections will be applied to two specific examples: the free particle and the harmonic oscillator. As before, we will work in euclidean time for simplicity. The aim is essentially to compute the function $\Pi(T)$. Although we have suggested a particular form for $A=A(T)$ in \eqref{epsilonA(T)}, we will leave $A$ as unknown, and only replace it at the end. 

\subsection{Free Particle: fixing the normalization} \label{Free Particle: fixing the normalization}
It is a general feature of the path integral method that one is usually able to compute physical quantities up to an overall normalization factor. Nevertheless the free particle is special in that one can determine uniquely its normalization by requiring that $\int_{-\infty}^\infty dx\ K(0,0;x,T)=1$, a calculation that only involves the classical action, without actually computing any path integral. But the method proposed in this paper only modifies the calculation of path integrals. Therefore as a heuristic approach, we will assume that the free particle propagator of the differentiable method matches the usual one, which is the well known
\begin{align}\label{}
K(x_1,0;x_2,T)=\left( \frac{m}{2\pi\hbar T} \right)^{1/2} e^{-\frac{1}{\hbar}S_{c}(x_1,x_2)}
\end{align}
for a particle going from $(x_1,0)$ to $(x_2,T)$, where $S_c$ is the (euclidean) action evaluated along the classical trajectory. Via path integrals, the same Kernel is, as in \eqref{PI0}
\begin{align}\label{}
K(x_1,0;x_2,T)=\mathcal{N} e^{-\frac{1}{\hbar}S_c(x_1,x_2)}\int_{0}^{0} \mathcal{D}x\ e^{-\frac{1}{\hbar} S[x]}
\end{align}
so equating these previous equations one obtains the normalization constant, which does depend on the space of paths being summed in the path integral:
\begin{align}\label{Norma}
\mathcal{N} \int_{0}^{0} \mathcal{D}x\ e^{-\frac{1}{\hbar} S[x]}=\left( \frac{m}{2\pi\hbar T} \right)^{1/2}
\end{align}

%%%%%%%%%%%%%%%%%%%%%%%%%%%%%%%%%%%%%%
\subsubsection{Feynman case}

In the usual Feynman method the amplitudes in the path integral are unbounded \eqref{FeynFour}, so we have
\begin{align}\label{intFey}
\int_{0}^{0} \mathcal{D}x\ e^{-\frac{1}{\hbar} S[x]}&= \prod_n \int_{-\infty}^{\infty} da_n\ e^{-\frac{mT}{4\hbar}\sum_n \lambda_n a_n^2} =\prod_n \left( \frac{4\pi\hbar }{mT\lambda_n} \right)^{1/2}
\end{align}
This fixes the Feynman normalization through \eqref{Norma} (the  functional determinant) which is standard textbook material:
\begin{align}\label{NFey}
\mathcal{N}_F=\left( \frac{m}{2\pi\hbar T} \right)^{1/2} \left( \prod_n \left( \frac{4\pi\hbar }{mT\lambda_n} \right)^{1/2} \right)^{-1}
\end{align}

%%%%%%%%%%%%%%%%%%%%%%%%%
\subsubsection{Free particle: differentiable case}

The PI in the differential case defined in (\ref{DiffFour}) drops out an extra Error function factor:
\begin{align}\label{PIDiff}
\int_{0\ D}^0 \mathcal{D}x\ e^{\frac{-1}{\hbar}S[x]}&=\left( \prod_n \int_{-A/n^\alpha}^{A/n^\alpha} da_n \right) e^{-\frac{T}{2}\frac{m}{2\hbar}\sum_n \lambda_n a_n^2}\\
&=\prod_n \left( \frac{4\pi\hbar}{mT \lambda_n } \right)^{1/2}\cdot \mbox{Erf} \left( \frac{A}{n^\alpha} \sqrt{\frac{mT\lambda_n }{4\hbar}}  \right) 
\end{align}
where the bounded gaussian integral \eqref{gaussianerror} was used. This fixes the modified ``differentiable" normalization, 
\begin{align}\label{NDiff}
\mathcal{N}_D%=\left( \frac{m}{2\pi\hbar T} \right)^{1/2} \left( \prod_n \left( \frac{4\pi\hbar}{mT \lambda_n } \right)^{1/2}\cdot \mbox{Erf} \left( \frac{A}{n^\alpha} \sqrt{\frac{mT\lambda_n }{4\hbar}}  \right)  \right)^{-1}\\
&=\mathcal{N}_F\left( \prod_n \mbox{Erf} \left( \frac{A}{n^\alpha} \sqrt{\frac{mT\lambda_n }{4\hbar}}  \right)  \right)^{-1}
\end{align}

In this heuristic approach, the differentiable method has no effect whatsoever upon the free particle (by construction), but it will have an important effect over any Lagrangian involving interactions. 

%%%%%%%%%%%%%%%%%%%%%%%%%%%%%%%55
\subsection{The Harmonic Oscillator}
%%%%%%%%%%%%%%%%%%%%%%%%%%%%%%%55

The harmonic oscillator (HO) serves as the primary example for any quantum computation. We start by rederiving the standard results for the usual PI by working in Euclidian space. Then the same logic will be applied for the PI over differentiable functions. 

\subsubsection{HO: Feynman case}

We briefly review here the standard derivation of the spectrum in the Feynman case. The Euclidian classical action for the harmonic oscillator gives rise to the kernel
\begin{align}\label{}
K_F(x_1,0;x_2, T)=e^{-\frac{1}{\hbar}S_c(x_1,x_2)}\mathcal{N}_F\int_F \mathcal{D}x\ e^{-\frac{1}{\hbar}\frac{m}{2}\int_0^T x(t)\left( -\partial_t^2+\omega^2 \right)x(t)}
\end{align}
which is exactly of the same kind of integral as computed in the free particle case (\ref{intFey}), but replacing $\lambda_n\rightarrow \lambda_n+\omega^2$. 
Using the Feynman normalization \eqref{NFey} and Gaussian integration one finds
%\begin{align}\label{}
%K_F(x_1,0;x_2, T)&=e^{-\frac{1}{\hbar}S_c(x_1,x_2)}\ \left( \frac{m}{2\pi \hbar T} \right)^{1/2} \frac{\prod_n \left( \frac{4\pi\hbar}{mT(\lambda_n+\omega^2)} \right)^{1/2}}{\prod_n \left( \frac{4\pi\hbar}{mT\lambda_n  } \right)^{1/2}} =e^{-\frac{1}{\hbar}S_c(x_1,x_2)}\ \left( \frac{m}{2\pi \hbar T} \right)^{1/2}  \prod_n \left( 1+\left( \frac{\omega T}{n\pi} \right)^2 \right)^{-1/2}\quad.
%\end{align}
%By further using Euler's solution to the ``Basel Problem" \eqref{EulerBasel}, this simplifies to
\begin{align}\label{}
K_F(x_1,0;x_2, T)&=e^{-\frac{1}{\hbar}S_c(x_1,x_2)}\ \sqrt{\frac{m}{2\pi \hbar }\cdot \frac{\omega}{\sinh(\omega T)}} \quad.
\end{align}
This Kernel allows to evaluate the 
partition function $Z$ as defined by the trace of the propagator, which can be written as a geometric series REF
\begin{align}\label{ZTE2}
Z_F(T)&=\int_{-\infty}^{\infty}dx\ K_F(x,0;x,T)\nonumber \\
&=\sum_{n=0}^\infty e^{-\omega (n+\frac{1}{2})T}\quad.
\end{align}
Finally, the energy levels are extracted from (\ref{ZTE2}) by interpreting it as a partition function
in the sense of statistical mechanics ($Z= \sum_n e^{-\beta E_n}$)
where $\beta=\frac{T}{\hbar}$ has the meaning of inverse temperature and
the summation goes over energy levels $E_n$ (recall $T$ stands for  time, not temperature).
Thus, the energy levels are obtained from a PI calculation:
\be
E_n=\hbar \omega\left( n+\frac{1}{2} \right)\quad.
\ee 

%%%%%%%%%%%%%%%%%%%%%%%%%%%%%%%55
\subsubsection{HO: Differentiable case}\label{subsubsec:HO: Differentiable case}

Following the same strategy as above, the first step in dealing with
the harmonic oscillator in the differentiable case consists in 
evaluating the Kernel.
By using the normalization \eqref{NDiff} and bounded Gaussian integrals
one obtains
\begin{align*}\label{}
K_D(x_1,0;x_2,T)&=e^{-\frac{1}{\hbar}S_c(x_1,x_2)}\cdot \mathcal{N}_D\cdot \int_D \mathcal{D}x\ e^{\frac{-1}{\hbar}\frac{m}{2}\int_0^Tx(t)\left( -\partial_t^2+\omega^2 \right)x(t)}\\ 
&=e^{-\frac{1}{\hbar}S_c(x_1,x_2)} \left( \frac{m}{2\pi \hbar T} \right)^{1/2} \frac{\prod_n \left( \frac{4\pi\hbar}{mT(\lambda_n+\omega^2)  } \right)^{1/2} \mbox{Erf} \left( \frac{A}{n^\alpha} \sqrt{\frac{mT(\lambda_n+\omega^2) }{4\hbar}}\right)}{\prod_n \left( \frac{4\pi\hbar}{m T\lambda_n } \right)^{1/2} \mbox{Erf} \left( \frac{A}{n^\alpha} \sqrt{\frac{m T\lambda_n }{4\hbar}}  \right) }
\end{align*}
which factorizes as the Feynman propagator times an infinite product
\begin{align}\label{}
K_D%=e^{-\frac{1}{\hbar}S_c}\ \left( \frac{m}{2\pi \hbar T} \right)^{1/2}   \left[  \prod_n \left( 1+\left( \frac{\omega T}{n\pi} \right)^2 \right) \right]^{-1/2}\cdot 
%\prod_n \frac{\mbox{Erf} \left( \frac{A}{n^\alpha} \sqrt{\frac{mT(\lambda_n+\omega^2) }{4\hbar}}\right)}{\mbox{Erf} \left( \frac{A}{n^\alpha} \sqrt{\frac{m T\lambda_n }{4\hbar}}  \right)}\\ \nonumber
&=K_{F}\cdot \prod_n \frac{\mbox{Erf} \left( \frac{A}{n^\alpha} \sqrt{\frac{mT(\lambda_n+\omega^2) }{4\hbar}}\right)}{\mbox{Erf} \left( \frac{A}{n^\alpha} \sqrt{\frac{m T\lambda_n }{4\hbar}}  \right)}
\end{align}

Therefore the function $\Pi(T)$ in this specific case is (in euclidean time) 
\begin{align}\label{Pi(B,tau)}
\Pi(T)&\equiv \prod_{n=1}^\infty \frac{\mbox{Erf} \left( \frac{B}{n^\alpha} \sqrt{((n\pi/T)^2+\omega^2)}\right)}{\mbox{Erf} \left( \frac{B}{n^\alpha} \sqrt{(n\pi/T)^2}  \right)}\hspace{1cm}B=A\sqrt{\frac{mT}{4\hbar}}
\end{align}
This infinite product incorporates all the effects of the differentiability condition
over the physics of the harmonic oscillator, such as the energy spectrum computed below. As in \eqref{ZTE2}, one continues on evaluating the D-partition
function by integrating the Kernel for identical initial and final positions, noting that the modification factor $\Pi$ is $x-$independent:
\begin{align}\label{ZDHO}
Z_D(T)&=\left( \int_{-\infty}^\infty dx\ K_{F}(x,0;x,T) \right) \Pi(T) %=\frac{1}{2\sinh(\omega T/2)}\cdot \Pi(T)
=Z_F(T)\cdot \Pi(T)
\end{align}
which again, factorizes as the Feynman partition function times the infinite product $\Pi(T)$. In order to extract the energy levels associated to $Z_D$, we replace \eqref{ZTE2} into \eqref{ZDHO}
\begin{align}\label{}
Z_D(T)%=\sum_n e^{-\frac{T}{\hbar}E_n} \cdot e^{\frac{T}{\hbar}\cdot \frac{\hbar}{T}\ln \left( \Pi(T) \right)}
=\sum_n e^{-\frac{T}{\hbar}\left( E_n-\frac{\hbar}{T}\ln(\Pi(T)) \right)}\equiv \sum_n e^{-\frac{T}{\hbar}E^D_n}
\end{align}
and therefore the modified energy levels $E^D_n$ are now time-dependent:
\begin{align}\label{E^D_n}
E_n^D(T)=E_n-\frac{\hbar}{T}\ln\left( \Pi(T) \right)
\end{align}
which matches exactly the conclusion arrived in \eqref{E_DSE}, when we extracted the energy spectrum from the modified Schrodinger equation, though the latter was written in real time. All levels are shifted by the same function, so the  spacing $\Delta E=E_{n+1}-E_n$ between successive levels remains unchanged. Experimentally, this means that this effect is not measurable in processes involving transitions between levels, for example. Nevertheless in order to have a definite experimental prediction, the actual function $\Pi(T)$ is needed, this depending on the particular choice of $A(T)$. Now, results for specific choices and their physical 
implications will be studied numerically.

%%%%%%%%%%%%%%%%%%%%%%%%%%%%%%%
\subsection{Numerics: Harmonic oscillator}\label{sec:NumRes}   

In subsection \ref{subsec: FreeDifferentiable case} we have argued that the most natural choice for $A=A(T)$ is \eqref{epsilonA(T)}:
\begin{align}\label{}
A(T)=\sqrt{\frac{\hbar T}{m}} \left( \frac{T}{\epsilon_D} \right)^{\alpha-1}
\end{align}

With this definite form  
of $A$, we now turn to compute $\Pi(T)$ from \eqref{Pi(B,tau)} (recall this is in imaginary time)
\begin{align}\label{Pie}
\Pi(T,\alpha)&\equiv \prod_{n=1}^\infty \frac{\mbox{Erf} \left[ \frac{\epsilon_D}{2} \left( \frac{T}{n \epsilon_D } \right)^{\alpha} \sqrt{((n\pi/T)^2+\omega^2)}\right]}{\mbox{Erf} \left[ \frac{\epsilon_D}{2} \left( \frac{T}{n \epsilon_D } \right)^{\alpha} \sqrt{(n\pi/T)^2}  \right] }
\end{align}

\begin{figure}[hbt]
   \centering
%\centerline{\protect\vbox{\epsfig{file=Approx.eps,
%width=0.6\textwidth}}}
\includegraphics[width=10cm]{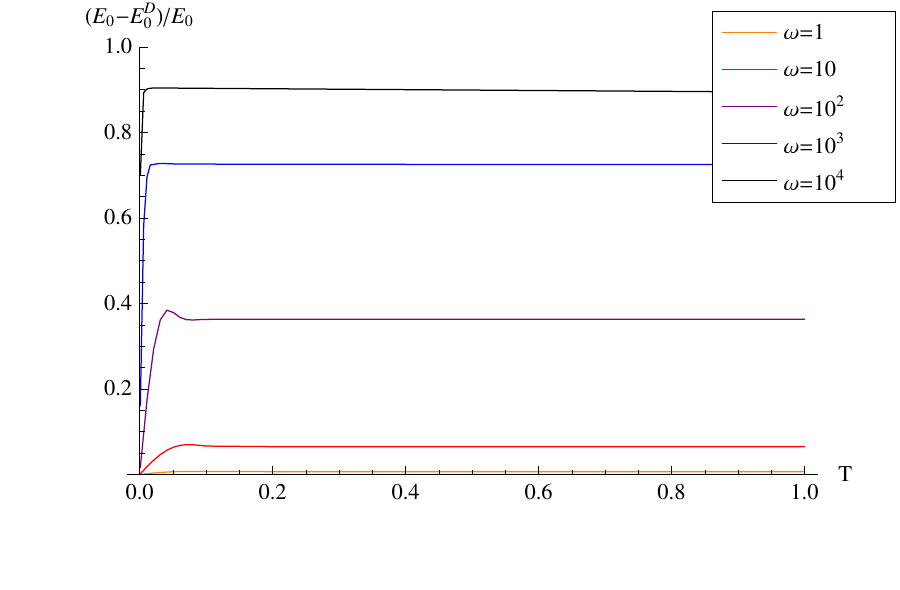}
  \caption{ Numerical calculation with $100.000$ terms, for $m=1,\hbar=1,\epsilon_D=0.1$, of the
  ground-state energy shift $E_0-E^D_0=\hbar\Delta \omega$ \eqref{E0DNum} over the conventional $E_0=\frac{\hbar \omega}{2}$, for $\alpha=2.1$ as function of time $T$ in the differentiable model, for increasing $\omega$.}
                \label{fig:dwEo}
\end{figure}

However, an analytic computation of infinite products such as this has remained inaccessible to the authors. Anyhow, numerical analysis of \eqref{Pie} is possible by considering a finite number of terms. In Figure \ref{fig:dwEo} we plot by performing numerical calculation of the harmonic oscillator energy levels \eqref{E^D_n} in the differentiable PI model:
\begin{align}\label{E0DNum}
E^D_0=\frac{\hbar \omega}{2}-\hbar\Delta \omega\hspace{1cm},\hspace{1cm} \Delta \omega =\frac{1}{T}\ln \left[ \Pi(T) \right]
\end{align}
Recall that it suffices to consider the ground state, since the energy shift $\Delta \omega$ is identical for every level.  
 \begin{figure}[hbt]
   \centering
%\centerline{\protect\vbox{\epsfig{file=Approx.eps,
%width=0.6\textwidth}}}
\includegraphics[width=10cm]{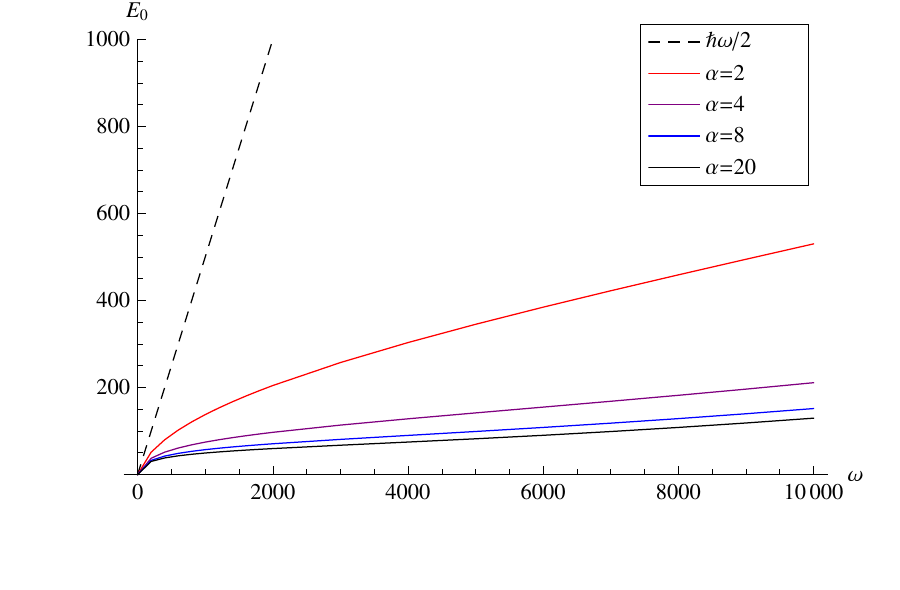}
  \caption{Numerical calculation with $n=100.000$ terms, for $m=1,\hbar=1,\epsilon_D=0.1$. 
  Absolute value of the ground-state energy as function of frequency $\omega$. In dashed the usual $\frac{\hbar \omega}{2}$; solid lines correspond to $E_0(\omega)$ in the differentiable model, for varying $\alpha$.}
                \label{fig:E0w}
\end{figure}
In Fig. \ref{fig:dwEo}, we see that:
\begin{itemize}
\item For $T\gtrsim \epsilon_D$ ($=0,1$ in this example), the energy shift $\Delta \omega= \frac{\ln\left[ \Pi(T) \right]}{T}$ is constant: $\Delta \omega$ is independent of time. Therefore the infinite product $\Pi(T)$ behaves as an exponential, $\Pi(T)=e^{\Delta \omega T}$, which rotated back to real time ($T\rightarrow iT$) means $\Pi(T)=e^{i\Delta \omega T}$. As explained in subsection \ref{subsec:The question of Unitarity} this is precisely the required form in order to respect unitarity.

\item For $T\lesssim \epsilon_D$, the modification is far from constant, so $\Pi(T)$ does not behave as an exponential. In \ref{subsec:The question of Unitarity} one sees demonstrated why this breaks unitarity. This was expected, since as it has been seen above that the laws of usual QM do not apply to time intervals shorter than $\epsilon_D$.

\item The percentage shift in the energy levels $\Delta \omega /\omega$ increases with $\omega$: for $\omega=1$ the shift is of order $\sim 1\%$, while for $\omega=10^4$ it is $\sim 90\%$. 
\end{itemize}
Fig. \ref{fig:E0w} presents the $\omega-$dependence of the various ground state energies
and allows for the following observations:
\begin{itemize}
\item The larger the differentiability exponent $\alpha$ is, the larger the deviation from $\frac{\hbar \omega}{2}$ is. 

\item However the curves seems to stabilize for $\alpha\rightarrow\infty$ 

\item The differentiable ground state energy for large $\omega$ behaves as $E^D_0(\omega)=a(\alpha)+b(\alpha)\omega$
\end{itemize}

With regard to the vacuum energy problem in QFT, a naive extrapolation of the last observation indicates that the vacuum density is lowered only by a few orders of magnitude, but is still highly divergent. Nevertheless it is interesting that a restriction over the space of paths can lead to a consistent quantum theory in which the ground state energy is strongly modified.

%%%%%%%%%%%%%%%%%%%%%%%%%%%%%%
\section{Discussion and summary} 
%%%%%%%%%%%%%%%%%%%%%%%%%%%%%%%

\subsection{Relation to other approaches}\label{sec:Relation to other approaches} 

Here are highlighted some other works found in the literature which may have some connection with the differentiable path integral approach put forward in this paper. 
\begin{enumerate}

\item \textbf{Higher-Derivatives Lagrangians.} One connection between the presented approach and previously studied models comes from higher-derivative theories, for example the well known Pais-Uhlenbeck oscillator \cite{Simon1990,BurakKovner2013}, which in principle corresponds to a ``perturbation" of a simple harmonic oscillator by a slightly modified mass term and a quadratic acceleration term:
\begin{align}\label{}
L&=\frac{1}{2}\left( 1+\lambda \omega^2  \right)\dot{x}^2-\frac{1}{2}\omega^2x^2-\frac{1}{2}\lambda^2\ddot x^2
\end{align} In essence by introducing higher-derivatives in the Lagrangian the action of the highly irregular paths becomes even more divergent than before, therefore suppressing more strongly their contribution to the path integral and thus rendering finite some quantities that were formerly divergent (for example $\langle v^2 \rangle $). However these theories present common problems as the appearance of ghosts, unitarity violation and energy spectrums not bounded from below, in this case
\begin{align}\label{}
E=(n+\frac{1}{2})\omega-(m+\frac{1}{2})\lambda^{-1}\ \ \ \ \mbox{for}\ \ n,m=0,1,2,\hdots
\end{align}
which is problematic since once would have expected to recover the simple harmonic oscillator as the perturbation goes to zero, $\lambda\rightarrow 0$, but instead gets an energy instability. As we have argued above, the model presented in this paper does not suffer from such effects.

\item \textbf{Causal Dynamical Triangulation.} It has also been proposed, specially in the context of quantum gravity, the possibility of summing only over space-time histories (paths) which always lie inside their local light cone \cite{IRWMS1992,Teitelboim1983,Gor2011,AGJL2013}. Some of them assume a discrete space-time in order to regularize the UV divergences. However it seems that, if one doesn't wish to rely upon discretizing space-time, any attempt of summing causally connected paths should be constructed on the basis of a differentiable path integral in which velocity makes sense locally (at least in the Lagrangian picture)

\item \textbf{Generalized Uncertainty Principle and Minimal Lenght.} 
Another connection may be established with the approach of Generalized Uncertainty Principle (GUP) and Minimal Length \cite{KMM1995, QuesneTkachuk2007,CLMT2011,DasPramanik2012}, as we already pointed out at \eqref{DGUP}. In recent years, there has been increasing interest in studying the possible existence of a minimal length scale motivated by string theory, loop quantum gravity, and non-commutative geometry.
Those approaches result in a modification of the propagator and of the canonical commutation relation \cite{Hossenfelder:2012jw}.
A popular realization \cite{Hossenfelder:2012jw} of a modified commutator takes the form
\begin{align}\label{GUP}
[x,p]=i\hbar \left( 1+\beta p^2 \right)\hspace{1cm}\beta>0\quad,
\end{align}
which is actually of the form that was found as approximation in this work.

Two main differences are evident compared with our result \eqref{DGUP}: 1) in the GUP/minimal-length context the relation \eqref{GUP} is frequently assumed to remain valid for arbitrarily high momentum, whereas in our model it is valid only for $p<p_D$ (a fixed value of momentum); 2) while GUP/minimal-length models frequently work with $\beta>0$, the differentiable model suggests $\beta<0$. 

\item \textbf{Maximal Acceleration.} The approach of Maximal Acceleration Hypothesis assumes the existence of an upper bound for the proper acceleration of massive particles \cite{Caianiello1982}. It has been shown to be closely related (via Quantum Geometry) to the GUP \cite{CLS2000}, and has been widely studied in High Energy Physics in relation to particle physics \cite{NFLS1999, Kuwata1996, CGP1988}, gravitation \cite{CGS1990, GS1989, CGS1991, GH1977}, and string theory \cite{McGuigan1994}. In this paper we have focused on a maximal velocity scenario (once differentiable paths), but a maximal acceleration would correspond to choosing $\alpha\geq 3$ (twice differentiable).

\end{enumerate}

%%%%%%%%%%%%%%%%%%%%%%%%%%%%%%%55
\subsection{Summary} \label{sec: Summary, Discussion and Conclusions}
%%%%%%%%%%%%%%%%%%%%%%%%%%%%%%%55

In this paper we have presented a method to restrict the space of paths entering the path integral for non-relativistic quantum mechanics, from Feynman's original space of ``all paths", to subspaces of $C^k$, therefore only allowing paths which possess at least $k$ derivatives. The method is most naturally implemented in Fourier space and introduces two external parameters, $A$ and $\alpha$, but we have focused mostly in the $C^1$ case. In terms of the geometry of sample paths, this replaces fractal/nowhere-differentiable trajectories (typical of Wiener processes) by functions which ``appear" to be fractal at time scales $\epsilon$ larger than $\epsilon_D$, but which are actually differentiable when examined at very small time scales $\epsilon<\epsilon_D$, where $\epsilon_D$ is the differentiable time scale which should be determined experimentally. This implies that the model behaves as usual QM at coarser scales, but looks very different for very short time intervals, and thus conceptually ``decouples" the high energy behavior of path integrals from its classical interpretation as a Brownian motion in imaginary time. \\

By computing the mean square velocity $\langle v^2 \rangle $, we found the usual $\sim \epsilon^{-1}$ dependence for $\epsilon>\epsilon_D$, while converging to a finite value for $\epsilon<\epsilon_D$, thus avoiding the otherwise divergent result. The canonical commutator becomes $[x,p]\approx \hbar (1-\beta p^2)$ for coarse scales ($\beta$ being very small parameter dependent on $\epsilon_D$ and $\alpha$), while it vanishes for $\epsilon<\epsilon_D$. The Schrodinger equation grabs an extra time-dependent potential whose net effect is to shift the energy eigenvalues of the system, leaving their spacing unchanged. A numerical analysis of the harmonic oscillator was presented, which suggests that the ground states energy $E_0(\omega)$ is strongly modified from $\frac{\hbar\omega}{2}$. Also, it was shown that an adequate choice of the parameter $A$ \eqref{epsilonA(T)} makes the evolution unitary which avoids the presence of ghosts.
In the present formulation it turns out that it generates
a shift in all the energy levels by the same amount, which is keeping the energy spacing between eigenstates unchanged. Thus it seems impossible to measure experimentally any effect involving instantaneous energy differences, e.g. atomic transitions. 
Therefore, one has to invoke an experiment involving the absolute value of energies, the first of which comes to mind is the Casimir effect. 
%Nonetheless it must be pointed out that in the last several years the question, wether or not the measured effect is actually a manifestation of the zero point energy, has been subject to discussion \cite{Jaffe2005}. Despite of this unresolved discussion, one can calculate the correction to the Casimir force by assuming that the measured effect actually comes due to the vacuum fluctuations.
%This allows to find an upper bound for the differentiable time scale $\epsilon_D$, which is of course only relevant if the vacuum interpretation of the Casimir effect is actually correct.
Such an estimate was realized in the appendix \ref{Casin} leading to the approximate prediction
that the differentiable time scale would have to be smaller than $\sim10^{-15}$~sec, if the Casimir would be 
a real vacuum effect. 

%Regarding the Heisenberg relation for very short times \eqref{dxdpD}, one may interpret in terms of time $\Delta t$ and energy $\Delta E$ uncertainties, in which case 
%\begin{align}\label{}
%\Delta E\ \Delta t \geq mv^2_{UV}\ \Delta t  \hspace{1cm}\mbox{for}\hspace{1cm}\Delta t<\epsilon_D
%\end{align}
%which suggests that there exist a limit on the energy that virtual particles can have for arbitrarily short times, in contrast with the usual $\Delta E\sim \frac{\hbar}{\Delta t}$ understanding. \\

The main results of this work is that, in contrast to the common believe, it is possible to
construct a consistent path integral quantum mechanics involving only differentiable paths, 
and that this construction further allows to render finite some
quantities that actually are divergent in the conventional approach to the PI formulation.

\begin{center}
\textbf{Awknowledgments}
\end{center} 
The work of B.K.\ was supported proj.\ Fondecyt 1120360
and anillo Atlas Andino 10201; the work of I.R by Conicyt-Pcha/MagNac/2012-22121934. The authors wish to acknoweldge to M. A. D\'iaz for his support throughout this project, and also M. Ba\~nados, M. Loewe, J. Mehringer and E. Mu\~noz for their helpful comments.

\pagebreak
%%%%%%%%%%%%%%%%%%%%%%%%%%%%%%%%%%%%%%%%%%%%%%%%%%%%%55
\begin{appendix}

%%%%%%%%%%%%%%%%%%%%%%%%%%%%%%%%%%%%%%

\section{Formulas and free particle computations}
\subsection{Some useful formulas}\label{Some useful formulas}

Gaussian integrals:
\begin{align}\label{gaussianerror}
\int_{-\infty}^\infty da\ e^{-ba^2}=\left( \frac{\pi}{b} \right)^{1/2}
\hspace{1cm},\hspace{1cm}
 \int_{-B}^B da\ e^{-ba^2} =\left( \frac{\pi}{b} \right)^{1/2}\mbox{Erf}\left( \sqrt{b}\ B \right) 
\end{align}
Taylor expansion of the Error function:
\begin{align}\label{}
\label{ErrorTay}
   \mbox{Erf}(z) = \left\{
     \begin{array}{lr}
            \frac{2z}{\sqrt{\pi}}+\mathcal{O}(z^2) & : z\ll 1\\
       1-\frac{e^{-z^2}}{\sqrt{\pi}z}+\mathcal{O}(e^{-z^2}z^{-2}) & : 1 \ll z
     \end{array}
   \right.
\end{align}

Very often we encounter the following quotient of integrals:
\begin{align}\label{}
\frac{\int_{-B}^B \xi^2 e^{-\alpha \xi^2} d\xi }{\int_{-B}^B e^{-\alpha \xi^2} d\xi }=\frac{1}{2\alpha}\left[ 1-\frac{2}{\sqrt{\pi}} \frac{\sqrt{\alpha}B e^{-\alpha B^2} }{\mbox{Erf}\left( \sqrt{\alpha}B \right)}  \right]\equiv \frac{1}{2\alpha}\left[ 1-Z(\sqrt{\alpha}B) \right]
\end{align}
which defines $Z(\cdot)$, whose Taylor expansions are:
\begin{displaymath}
   1-Z(W) = \left\{
     \begin{array}{lr}
            1-\frac{2}{\sqrt{\pi}}\sqrt{W}e^{-W} & : 1 \ll W\\
       \frac{2W}{3} & : W \ll 1
     \end{array}
   \right.
\end{displaymath} 

%Euler's formula for the infinite product (and its analytic continuation):
%\begin{align}\label{EulerBasel}
%\prod_{n=1}^\infty \left( 1-\left( \frac{x}{n\pi} \right)^2 \right)=\frac{\sin(x)}{x}\ \ \Rightarrow\ \ \prod_{n=1}^\infty \left( 1+\left( \frac{x}{n\pi} \right)^2 \right)=\frac{\sin(ix)}{ix}=\frac{\sinh(x)}{x}
%\end{align}

\subsection{Computation of Feynman Series} \label{Feynman Series}
Rewrite the series in \eqref{v2Fseries} as sum of exponentials first, calling $\frac{i\pi t_0}{T}\equiv \mu_0$ and $\frac{i\pi\epsilon}{T}\equiv \mu$
\begin{align*}\label{}
S_F&=\sum_j \frac{1}{j^2} \left[ \sin \left( \frac{j\pi (t_0+\epsilon)}{T} \right) - \sin \left( \frac{j\pi t_0}{T} \right) \right]^2=\sum_j \frac{1}{j^2} \left[ \frac{1}{2i}\left( e^{j(\mu_0+\mu)}-e^{-j(\mu_0+\mu)} \right)-\frac{1}{2i}\left( e^{j\mu_0}-e^{-j\mu_0} \right) \right]^2\\
%&=-\frac{1}{4}\sum_j \frac{1}{j^2} \left[     \left( e^{j(\mu_0+\mu)}-e^{-j(\mu_0+\mu)} \right)^2-2\left( e^{j(\mu_0+\mu)}-e^{-j(\mu_0+\mu)} \right) \left( e^{j\mu_0}-e^{-j\mu_0} \right) +\left( e^{j\mu_0}-e^{-j\mu_0} \right)^2   \right]\\
&=-\frac{1}{4}\sum_j \frac{1}{j^2} \left[ e^{2j(\mu_0+\mu)}-2+ e^{-2j(\mu_0+\mu)} -2\left( e^{(2j\mu_0+j\mu)}-e^{j\mu}-e^{-j\mu}+e^{-(2j\mu_0+j\mu)} \right)+e^{2j\mu_0}-2+e^{-2j\mu_0} \right]\\
&=\zeta(2)-\frac{1}{4}\sum_j \frac{1}{j^2} \left[ e^{2j(\mu_0+\mu)} -2\left( e^{(2j\mu_0+j\mu)}-e^{j\mu}\right)+e^{2j\mu_0} \right]-\frac{1}{4}C.C.
\end{align*}
where we have extracted the constant term $\sum_j \frac{1}{j^2}=\zeta(2)=\frac{\pi^2}{6}$, and $C.C$. indicates the complex conjugate of the former series, which now takes the precise form of a dilogarithm (or more properly, its analytic continuation)
\begin{align}\label{}
Li_2(z)=\sum_{j=0}^\infty \frac{z^j}{j^2}
\end{align}
and thus
\begin{align}\label{A2}
S_F&=\zeta(2)-\frac{1}{4}\left[ Li_2\left( e^{2(\mu_0+\mu)}\right) -2Li_2\left( e^{2\mu_0+\mu} \right)  +2Li_2\left( e^{\mu} \right)+Li_2(e^{2\mu_0})  \right]-\frac{1}{4}C.C.
\end{align}
Now, our expansion consists on making $\mu\sim \frac{\epsilon}{T}\rightarrow 0$ while $\mu_0$ remains constant. As the Dilogarithm is analytical there, we can expand around $\mu_0$, and use that its derivatives are
\begin{align}\label{}
\frac{\partial Li_s (e^{\mu})}{\partial \mu}=Li_{s-1}(e^\mu)
\end{align}
and therefore, to first order in $\mu\approx 0$:
\begin{align*}\label{}
\bullet\ \ \ &Li_2\left( e^{2(\mu_0+\mu)} \right)=Li_2\left( e^{2\mu_0} \right)+Li_1\left( e^{2\mu_0} \right)\cdot 2\mu+\mathcal{O}(\mu^2)\\
\bullet\ \ \ &Li_2\left( e^{2\mu_0+\mu} \right)=Li_2\left( e^{2\mu_0} \right)+Li_1\left( e^{2\mu_0} \right)\cdot \mu+\mathcal{O}(\mu^2)
\end{align*}
and replacing into \eqref{A2}, many factors cancel out, leaving
\begin{align*}\label{}
S_F&=\zeta(2)-\frac{1}{4}\Big[ Li_2\left( e^{2\mu_0} \right)+Li_1\left( e^{2\mu_0} \right)\cdot 2\mu -2 \left( Li_2\left( e^{2\mu_0} \right)+Li_1\left( e^{2\mu_0} \right)\cdot \mu \right)   +2Li_2\left( e^{\mu} \right)+Li_2(e^{2\mu_0})  \Big]-\frac{1}{4}C.C.+\mathcal{O}(\mu^2)\\
&=\zeta(2)-\frac{1}{2} Li_2\left( e^{\mu} \right) -\left( \frac{1}{2} Li_2\left( e^{\mu} \right) \right)^*+\mathcal{O}(\mu^2)
\end{align*}

Finally, we must expand the remaining dilogarithm; for this we use the expression \cite{Wood1992,GradRyz1980} (valid for $|\mu|<2\pi$)
\begin{align}\label{}
Li_s(e^\mu)=\frac{\mu^{s-1}}{(s-1)!}\left[ H_{s-1}-\ln(-\mu) \right]+\sum_{k=0,k\neq s-1}^\infty \frac{\zeta(s-k)}{k!}\mu^k\hspace{1cm}H_s=\sum_{h=1}^s \frac{1}{h}
\end{align}
by which, using $s=2$ and $\mu\approx 0$ (note that $H_1=1$ and $k\neq s-1=1$)
\begin{align}\label{}
Li_2(e^\mu)\approx \mu\left[ 1-\ln(-\mu) \right]+\zeta(2)+\mathcal{O}(\mu^2)
\end{align}
and thus, noting that conjugation means $\mu=\frac{i\pi\epsilon}{T}\rightarrow -\mu$, we arrive at:
\begin{align}\label{}
S_F&=\zeta(2)-\frac{1}{2} \left[ Li_2\left( e^{\mu} \right)+Li_2\left( e^{-\mu} \right) \right] +\mathcal{O}(\mu^2)
\end{align}
and furthermore
\begin{align}\label{}
Li_2(e^\mu)+Li_2(e^{-\mu})&\approx \mu\left[ 1-\ln(-\mu) \right]+\zeta(2) -\mu\left[ 1-\ln(\mu) \right]+\zeta(2)+\mathcal{O}(\mu^2)\\
%&=\mu\left[ \ln(\mu)-\ln(-\mu) \right]+2\zeta(2)+\zeta(0)\mu^2+\mathcal{O}(\mu^2)\\
%&=\mu\ln(-1)+2\zeta(2)+\mathcal{O}(\mu^2)\\
&=i\pi\mu+2\zeta(2)+\mathcal{O}(\mu^2)
\end{align}
where we replaced $\ln(-1)=i\pi,\zeta(0)=-\frac{1}{2}$. Therefore, replacing $\mu=\frac{i\pi\epsilon}{T}$ we finally arrive at
\begin{align}\label{}
S_F(\epsilon)&=\zeta(2)-\frac{1}{2}\left[ i\pi\mu+2\zeta(2) \right] +\mathcal{O}(\mu^2)=\left[  \frac{\pi^2}{2}\frac{\epsilon}{T}   \right] +\mathcal{O}\left( \left( \frac{\epsilon}{T} \right)^2 \right)\label{SFcomputed}
\end{align}

Note that $S_F(\epsilon)$ is  continuos: it converges since it has $\alpha=2$. And since $S_F(0)=0$, we confirm that it cannot have a constant contribution independent of $\epsilon$.

\subsection{Computation of DPI Series} \label{Differentiable Series}
Now we turn to compute the sum \eqref{55}, keeping in mind the case $\alpha>2$ so that all series converge. We are not interested in exact numerical values, but rather only on the dependence of the series upon the parameters $A$ and $\alpha$, and therefore we will only seek for upper bounds as means of estimating the various functions involved. Also recall from \eqref{barA} that
\begin{align}\label{barAap}
\bar{A}=\sqrt{\frac{m \pi^2}{4\hbar T}  } A
\end{align}

\newpage
\subparagraph{``Differentiable"/High resolution region $\epsilon<\epsilon_D$}\ \\

Due to the complexity of the modified series, we won't compute exact values, but rather find bounds on it such as to have a feeling on its dependence on the various parameters. The sum to analyze is the one in \eqref{55}, which we will call $S_D(\tau)$, where $\tau=\frac{\pi\epsilon}{T}$ is the resolution scale (recall also \eqref{Z y W QM} and \eqref{barA})
\begin{align}\label{SDtau}
S_D(\tau)=\sum_{j=1}^\infty \frac{\sin^2(j\tau)}{j^2}\left( 1-Z_j \right)=\sum_{j=1}^\infty\frac{\sin^2(j\tau)}{j^2}\left( 1-\frac{2}{\sqrt{\pi}}\frac{\sqrt{\frac{\bar{A}}{j^{\alpha-1}}}e^{-\bar{A}/j^{\alpha-1}}}{\mbox{Erf}\left( \sqrt{\bar{A}/j^{\alpha-1}} \right)} \right)
\end{align}
and consider expressing it as a Taylor series around $\tau=0$: 
\begin{align}\label{}
S_D(\tau)=S_D(0)+S'(0)\cdot \tau+\frac{1}{2}S''(0)\cdot \tau^2+\mathcal{O}(\tau^3)
\end{align}

First note that $S_D(\tau)$ is continuos and differentiable, since as $j\rightarrow \infty$ (using the Error's expansion \eqref{ErrorTay})
\begin{align}\label{}
\frac{\sqrt{\frac{\bar{A}}{j^{\alpha-1}}}}{\mbox{Erf}\left( \sqrt{\bar{A}/j^{\alpha-1}} \right)}\approx %\frac{\sqrt{\pi}}{2}+\frac{\sqrt{\pi} \sqrt{\frac{\bar{A}}{j^{\alpha-1}}} }{6}\sqrt{\bar{A}/j^{\alpha-1}}+\mathcal{O}(\bar{A}/j^{\alpha-1})^{3/2}
\frac{\sqrt{\pi}}{2}+\frac{\sqrt{\pi}  }{6}\cdot \frac{\bar{A}}{j^{\alpha-1}}+\mathcal{O}\left(\frac{\bar{A}}{j^{\alpha-1}}\right)^{3/2}
\end{align}
so the addends inside the sum decays like $\sim j^{-2-(\alpha-1)}$, thus ensuring the convergence of the sum and its derivative (if $\alpha>2$). Evaluating \eqref{SDtau} at $\tau=0$ gives us $S_D(0)=0$, and evaluating its derivative at $\tau=0$ we also get $S'(0)=0$, and therefore $S_D(\tau)$ is quadratic at lowest order:
\begin{align}\label{SDTAU}
S_D(\tau)=\frac{1}{2}S''(0)\cdot \tau^2+\mathcal{O}(\tau^3)
\end{align}

Next we compute the second derivative of \eqref{SDtau}, which is:
\begin{align}\label{}
S''_D(\tau)=2\sum_j \cos(2j\tau)\left( 1-Z_j \right)\ \ \Rightarrow\ S''_D(0)= 2\sum_j \left( 1-Z_j \right)
\end{align}
Since the Error function inside $Z_j$ in \eqref{SDtau} is difficult to manage, we will resort to using the following bound property (which is deduced from its Taylor expansion)
\begin{align}\label{}
1-Z_j\leq 
\left\{
     \begin{array}{lr}
     1-\frac{2}{\sqrt{\pi}} \sqrt{\frac{\bar{A}}{j^{\alpha-1}}}  & :\ \ \ j < \bar{A} \\
      \\
      \frac{2}{3}\frac{\bar{A}}{j^{\alpha-1}} & :\ \ \ \bar{A} <j \\
     \end{array}
   \right.
\end{align}

We start by splitting the sum in two according to these upper bounds:
\begin{align}\label{}
\frac{1}{2}S_D''(0)&= \sum_{j=1}^\infty \left( 1-Z_j \right)=\sum_{j=1}^{\bar{A}} \left( 1-Z_j \right)+\sum_{j=\bar{A}}^\infty \left( 1-Z_j \right)\\
&\leq  \underbrace{\sum_{j=1}^{\bar{A}}  \left( 1-\frac{2}{\sqrt{\pi}}  \sqrt{\frac{\bar{A}}{j^{\alpha-1}}}e^{-\bar{A}/j^{\alpha-1}} \right)}_{\mbox{first sum}}+\underbrace{\sum_{j=\bar{A}}^\infty  \frac{2\bar{A}}{3j^{\alpha-1}}}_{\mbox{second sum}}
\end{align}
Both sums can now be bounded from above using the integral criterion. A direct computation of this integral gives, for any $\alpha>2$, that the leading order as $A\rightarrow \infty$ is simply
\begin{align*}\label{}
\mbox{first sum}&\leq \int_1^{\bar{A}} dj\left( 1-\frac{2}{\sqrt{\pi}}  \sqrt{\frac{{\bar{A}}}{j^{\alpha-1}}}e^{-{\bar{A}}/j^{\alpha-1}} \right)\leq \bar{A}  %  -1+{\bar{A}}\left( \underbrace{1-\frac{4}{e\sqrt{\pi}}-4\mbox{Erf}(1)+4\frac{e^{-{\bar{A}}}}{\sqrt{{\bar{A}}}\sqrt{\pi}}+4\mbox{Erf}(\sqrt{{\bar{A}}})}_{\gamma_1} \right)\leq \gamma_1 {\bar{A}}
\end{align*}
%where $\gamma_1$ is a constant that may be defined in the limit $A\rightarrow \infty$,  $\gamma_1\equiv 5-\frac{4}{e\sqrt{\pi}}-4\mbox{Erf}(1)\approx 0,798... \mbox{for}\ \ \alpha=2$ since we expect to take $A\gg 1$.
 Now for the second sum
\begin{align}\label{}
\sum_{j={\bar{A}}}^\infty  \frac{1}{j^{\alpha-1}}\leq \int_{\bar{A}}^\infty \frac{dj}{j^{\alpha-1}}=\frac{1}{\alpha-2}\cdot\frac{1}{\bar{A}^{\alpha-2}}%=\frac{1}{\eta}\cdot \frac{1}{A^{\eta}}
\end{align}
which again reflects the fact that the velocity is divergent $\alpha\leq 2$ (paths are fractal). Of course, we wish to consider the region where $\alpha>2$ of differentiable paths. 
Thus we arrive at a upper bound for the desired term in \eqref{SDTAU}
\begin{align}\label{}
\frac{1}{2}S''(0)&\leq {\bar{A}}+\frac{2}{3}\frac{\bar{A}^{3-\alpha}}{(\alpha-2)}%\ \ \ \Rightarrow\ \ \ \frac{1}{2}S''(0)\leq \gamma_1 {\bar{A}}+\frac{2}{3}\frac{1}{\eta}\cdot\frac{1}{{\bar{A}}^{\eta-1}}
\end{align}
%when expanded above the critical value $\alpha=2+\eta$, and where $\gamma_1$ is a finite constant which depends on $\alpha$ (here computed considered as $2$), although its dependence is probably inconsequential for our purposes. 
but recall we are considering $\alpha>2$, so unless we take $\alpha\rightarrow 2^+$ in general the first term dominates over the second as $A\rightarrow \infty$, so for the differentiable region $\tau<\tau_D$ we find
\begin{align}\label{}
 S_D(\tau)\leq \bar{A}\ \tau^2  \hspace{.5cm}\mbox{for}\hspace{.5cm}\tau<\tau_D
\end{align}

Finally, this translates via \eqref{55} into the mean square velocity for the differentiable region, replacing \eqref{barAap}:
\begin{align}\label{v2Dhigh}
\langle v^2 \rangle _D(\epsilon)=\frac{2\hbar}{mT} \frac{1}{\tau^2} S_D(\tau) = \frac{2\hbar}{mT} \bar{A}= \frac{\pi A}{T}\cdot \sqrt{\frac{\hbar}{m T}} \equiv v^2_{UV} \hspace{.5cm}\mbox{for}\hspace{.5cm}\epsilon<\epsilon_D 
\end{align}

Interestingly enough, we see that the differentiable UV-velocity squared is the product of a ``quantum-like" velocity $\sqrt{\frac{\hbar}{ mT}}$ in the line of \eqref{v^2Fey}, and a ``classical-like" velocity $\frac{A}{T}$, which is of course the only possible combination involving the external parameter of the length $A$.

\subparagraph{``Feynman"/Low resolution region $\epsilon_D<\epsilon$}\ \\

As seen in Figure \ref{fig:v2}, when $\tau_D<\tau$ both the Feynman and the differentiable curves are very close, so we may treat the modification factor, here $Z_j$, as a perturbation:
\begin{align}\label{SD=SF-C}
S_D(\tau)=\sum_{j=1}^\infty \frac{\sin^2(j\tau)}{j^2}\left( 1-Z_j \right)=S_F(\tau)-\sum_{j=1}^\infty \frac{\sin^2(j\tau)}{j^2}Z_j
\end{align}
where $S_F(\tau)=\frac{\pi}{2}\tau$ is the Feynman sum (see \ref{SFcomputed}). So all that remains is to compute this last series: 
\begin{align}\label{Csum}
\sum_{j=1}^\infty \frac{\sin^2(j\tau)}{j^2}Z_j=\sum_{j=1}^\infty \frac{\sin^2(j\tau)}{j^2}\frac{2}{\sqrt{\pi}} \frac{\sqrt{\bar{A}/j^{\alpha-1}}\ e^{-\bar{A}/j^{\alpha-1}} }{\mbox{Erf}\left( \sqrt{\bar{A}/j^{\alpha-1}} \right)}
\end{align}

We will seek again for an upper bound for this series. Once more we separate the sum as
\begin{align}\label{sumfirstsecond}
\sum_{j=1}^\infty \frac{\sin^2(j\tau)}{j^2}Z_j=\underbrace{\sum_{j=1}^{\bar{A}} \frac{\sin^2(j\tau)}{j^2}Z_j}_{\mbox{first}}+\underbrace{\sum_{j={\bar{A}}}^\infty\frac{\sin^2(j\tau)}{j^2}Z_j}_{\mbox{second}}
\end{align}
Let's begin by finding an upper bound for the second sum in \eqref{sumfirstsecond}, using that $e^{-{\bar{A}}/j^{\alpha-1}}\leq 1$
\begin{align}\label{}
\mbox{second}\leq \frac{2}{\sqrt{\pi}} \sqrt{{\bar{A}}} \sum_{j={\bar{A}}}^\infty \frac{1}{j^{(3+\alpha)/2}} \frac{ 1 }{\mbox{Erf}\left( \sqrt{\bar{A}/j^{\alpha-1}} \right)}
\end{align}
In this region, $x=\bar{A}/j^{\alpha-1}\rightarrow 0$ as $j\rightarrow \infty$, so using the Taylor expansion of the inverse error function,
%\begin{align}\label{}
%\frac{1}{\mbox{Erf}(x)}=\frac{\sqrt{\pi}}{2} \frac{1}{x}+\mathcal{O}(x^3)
%\end{align}
%and as $x\rightarrow 0$, 
we will estimate the sum as (the exact value is irrelevant)
\begin{align}\label{second}
\mbox{second}\leq \frac{2}{\sqrt{\pi}} \sqrt{{\bar{A}}} \sum_{j={\bar{A}}}^{\infty} \frac{1}{j^{(3+\alpha)/2}} \frac{\sqrt{\pi}}{2}  \sqrt{\frac{j^{\alpha-1}}{{\bar{A}}}}\leq \int_{\bar{A}}^\infty \frac{dj}{j^{2}}=\frac{1}{{\bar{A}}}
\end{align}
independently of the value for $\alpha$. Next for the first sum, we use that $\sin^2(j\delta)<1$, and we can replace $\frac{1}{\mbox{Erf}(\sqrt{\bar{A}/j^{\alpha-1}})}$ by its maximum value (within $1<j\leq \bar{A}$) which is $\frac{1}{\mbox{Erf}(1)}$ and then apply the integral criterion, retaining the leading order as $A\rightarrow \infty$:
\begin{align}\label{first}
\mbox{first}\leq  \frac{2}{\sqrt{\pi}}\sqrt{{\bar{A}}} \sum_1^{\bar{A}} \frac{e^{-\bar{A}/j^{\alpha-1}}}{j^{(3+\alpha)/2}} \leq  \frac{2}{\sqrt{\pi}}\frac{\sqrt{{\bar{A}}}}{\mbox{Erf}(1)} \int_{1}^{\bar{A}} dj\frac{e^{-\bar{A}/j^{\alpha-1}}}{j^{(3+\alpha)/2}} \sim \frac{1}{\bar{A}^{\alpha-1}}       
%= \frac{1}{{\bar{A}}^{\frac{1}{\alpha-1}} }f(\eta,{\bar{A}}) +\mathcal{O}(e^{-{\bar{A}}})
\end{align}
but as we are assuming $\alpha>2$, \eqref{first} will always decrease faster than \eqref{second} as $A\rightarrow \infty$, so we find that \eqref{Csum} is dominated by \eqref{second}:
\begin{align}\label{}
\sum_{j=1}^\infty \frac{\sin^2(j\tau)}{j^2}Z_j \ \lesssim\ \frac{1}{\bar{A}} \hspace{.5cm}\mbox{for}\hspace{.5cm} \tau_D<\tau
\end{align}
that through \eqref{55} and using \eqref{SD=SF-C} implies a velocity for the low resolution region $\epsilon_D<\epsilon$ of
\begin{align}\label{v2Dlowfinal}
\langle v^2 \rangle _D(\epsilon)&%=\frac{2\hbar}{mT} \frac{1}{\tau^2} S_D(\tau)
=\frac{2\hbar}{mT} \frac{1}{\tau^2} \left( S_F(\tau)-\frac{1}{\bar{A}} \right)%=\langle v^2 \rangle _F-\frac{2\hbar}{mT \bar{A}} \frac{1}{\tau^2}
=\langle v^2 \rangle _F-\frac{2\hbar T}{\pi^2 m \bar{A}} \frac{1}{\epsilon^2}=\langle v^2 \rangle _F-\frac{4}{\pi^3}\frac{1}{A}\left( \frac{\hbar T}{m} \right)^{3/2}\frac{1}{\epsilon^2}
%&=\langle v^2 \rangle _F-\frac{2\hbar}{mT} \frac{\gamma_2}{\pi^2\epsilon^2}\frac{T^2}{A}
\end{align}
where we have replaced \eqref{barAap}.

%\subparagraph{Differentiable model: summary on velocities}\ \\

%Finally, taking into account \eqref{v2Dhigh} and \eqref{v2Dlowfinal}, and replacing \eqref{barAap} we have in brief
%\begin{align}\label{}
%   \langle v^2 \rangle _D = \left\{
%     \begin{array}{lr}
%     v^2_{UV}  \hspace{1.5cm} \epsilon <\epsilon_D \hspace{1.5cm}v^2_{UV}=\frac{\pi A}{T}\cdot \sqrt{\frac{\hbar}{m T}} \\
%      \\
%       \frac{\hbar}{m\epsilon}-\frac{C}{\epsilon^2}  \hspace{1cm} \epsilon_D<\epsilon \hspace{1.5cm}C=\frac{4}{\pi^3}\frac{1}{A}\left( \frac{\hbar T}{m} \right)^{3/2} 
%     \end{array}
%   \right. 
%\end{align} 
%
%%----------------------------
%%%
%%%

%%%%%%%%%%%%%%%%%
\section{Casimir effect: one dimensional toy model} \label{Casin}

 The $d=3+1$ computation in the differentiable model is rather arduous technically, so we work in the simplified $d=1+1$ scenario (although this has not been measured experimentally). Briefly, the relevant quantity to compute for the Casimir effect is the energy difference between the assembly with the two plates (with null boundary conditions for the electric field) and the system without them (no boundary conditions). In usual QM, calling $L$ the distance between the two (one-dimensional) plates, the allowed wavelengths between plates are $\lambda_n=\frac{2L}{n}$ which implies the allowed energies
\begin{align}\label{E_n(L)}
E_{n}(L)=\frac{1}{2}\hbar \omega_n  = \frac{1}{2}\frac{ \hbar c \pi  n }{L}
\end{align} 
and with this, 
 \begin{align}\label{dECasimir}
\Delta E(L)=E(L)-E(\infty)= \frac{1}{2} \frac{ \hbar c \pi }{L} \left( \sum_{n=0}^\infty n - \int_{0}^\infty n\ dn  \right)\hspace{.5cm}\Rightarrow\hspace{.5cm} = \frac{1}{2}\frac{ \hbar c \pi }{L} \left( \sum_{n=0}^\infty n\ g(n/n_c) - \int_{0}^\infty n\ g(n/n_c) \ dn  \right)
\end{align}
where one is compelled to introduce some smooth regularizating/cut-off function $g(n/n_c)$ such that $g=1$ for $n\ll n_c$ and $g=0$ for $n_c\ll n$, its specific form being irrelevant for the present purpose. Once \eqref{dECasimir} is finite, one may use the Euler-Maclaurin formula which reads, for any $f(n)$ 
\begin{align}\label{Euler-Maclaurin}
\sum_{n=0}^\infty f(n) - \int_{0}^\infty f(n) \ dn = -\sum_{k=1}^\infty \frac{B_k}{k!} f^{(k-1)}(0)
\end{align}
where $f$ is assumed to be regularized, and $B_k$ are the Bernoulli numbers. In usual QM, $f(n)=n$ and \eqref{Euler-Maclaurin} gives 
\begin{align}\label{dEusual}
\Delta E(L)&=\frac{1}{2}\frac{\hbar c\pi }{L} \left( -\frac{1}{12} \right)
\end{align}

\begin{figure}[hbt]
   \centering
%\centerline{\protect\vbox{\epsfig{file=Approx.eps,
%width=0.6\textwidth}}}
\includegraphics[width=10cm]{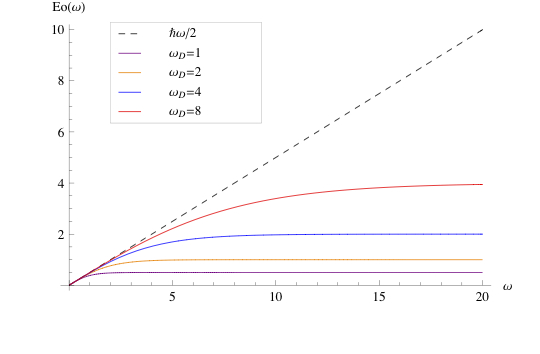}
\caption{Toy model for ground state, $E_D(\omega)=\frac{1}{2}\hbar \omega_D \tanh\left( \frac{\omega}{\omega_D} \right)$}
\label{fig:tanh}
\end{figure}

In the differentiable case, we don't have an explicit analytic expression like \eqref{E_n(L)} for $E(\omega)$. One can use however the numerical analysis from the previous section. Therefore, and based on the numerical results from \ref{sec:NumRes} (in particular Fig.4b), we will take as a toy model curve for the HO ground state energy in the differentiable scenario the following:
\begin{align}\label{E_Dtoy}
E_D(\omega)=\frac{1}{2}\hbar \omega_D \tanh\left( \frac{\omega}{\omega_D} \right)
\end{align}
where $\omega_D$ is a regulable frequency parameter; \eqref{E_Dtoy} is depicted in Fig. \ref{fig:tanh} for different values of $\omega_D$. The same boundary conditions of above apply also here, and in order to compare to the conventional result \eqref{dECasimir} we define the analog of $f$ in \eqref{Euler-Maclaurin} as $f_D$ here by
\begin{align}\label{}
E_D(\omega_n)=\frac{1}{2}\frac{ \hbar c \pi }{L} \cdot \frac{L}{  c \pi }\omega_D \tanh\left( \frac{\omega_n}{\omega_D} \right)\equiv \frac{1}{2}\frac{ \hbar c \pi }{L} \cdot f_D(n)
\end{align}
where 
\begin{align}\label{}
f_D(n)=\frac{L\omega_D}{c\pi} \tanh\left( \frac{c\pi}{L\omega_D} n \right)
\end{align}

Therefore, the RHS of the Euler-Maclaurin formula \eqref{Euler-Maclaurin} will be in this case ($B_2=\frac{1}{6}, B_4=-\frac{1}{30}$)
\begin{align}\label{}
\sum_{k=1}^\infty \frac{B_k}{k!} f_D^{(k-1)}(0)&=\sum_{k=1}^\infty \frac{B_k}{k!} \left( \frac{\pi c}{L\omega_D} \right)^{k-2} \frac{d^{(k-1)}}{dn^{(k-1)}}\tanh\left( n \right)|_{n=0}\\
&=-\frac{1}{12}-\frac{\frac{2}{30}}{4!}\left( \frac{\pi c}{L\omega_D} \right)^2 +\mathcal{O}\left( \frac{\pi c}{L\omega_D} \right)^4
\end{align}

This amounts to an energy difference of
\begin{align}\label{dE(L)D}
\Delta E(L)=\frac{1}{2}\hbar \frac{\pi c}{L} \left[ -\frac{1}{12}-\frac{1}{40}\left( \frac{\pi c}{L\omega_D} \right)^2 +\mathcal{O}\left( \frac{\pi c}{L\omega_D} \right)^4 \right]
\end{align}
which is to be contrasted with the usual \eqref{dEusual}. Clearly we see that the corrections come as powers of $\frac{\pi c}{L\omega_D}$, which is to be regarded as the ``smallness" parameter. If we take $L_{\mbox{exp}}\sim 0,1\mu m$ as the shortest distance at which Casimir's effect has been measured (although that is valid for $d=3+1$) with an error of $\sim1\%$ \cite{Lam1997,MR1998}, then by demanding that the correction term in \eqref{dE(L)D} is lesser than $1\%$ of the first term gives a lower bound for $\omega_D$ in order of magnitude of
\begin{align}\label{}
\omega_D>\frac{c}{L_{\mbox{exp}}}
\end{align}

Finally, since it must occur that $\omega_D\rightarrow\infty$ as $\epsilon_D\rightarrow 0$, and also by dimensional analysis, one may expect that $\omega_D\sim \epsilon_D^{-1}$ and thus we find an upper bound for the order of magnitude of the differentiable time scale $\epsilon_D$
\begin{align}\label{}
\epsilon_D<\frac{L_{\mbox{exp}}}{c}\sim 10^{-15}s
\end{align}
\end{appendix}

%%%%%%%%%%%%%%%%%%%%%%%%%%%%%%%%%%%%%%%%%%%%%%%%%%%%%55

%\end{thebibliography}

\end{document}